\documentclass[onecolumn,showpacs,showkeys,letterpaper]{revtex4}

\usepackage{amsmath}    
\usepackage{amstext}    
\usepackage{amssymb}    
\usepackage{amsfonts}   
\usepackage{graphicx}   
\usepackage{color}     
\usepackage{hyperref}   
\newcommand{\C}[2]{\left(\!\begin{array}{c}#1\\#2\end{array}\!\right)}
\renewcommand{\O}[1]{\mathcal{O}\left(#1\right)}

\begin{document}

\title{Cumulants of the current in the weakly asymmetric exclusion process}
\author{Sylvain Prolhac, Kirone Mallick}
\affiliation{Institut de Physique Th\'eorique\\CEA, IPhT, F-91191 Gif-sur-Yvette, France\\CNRS, URA 2306, F-91191 Gif-sur-Yvette, France}
\date{March 11, 2009}

\begin{abstract}
We study the fluctuations of the total current for the partially asymmetric exclusion process in the scaling of a weak asymmetry (asymmetry of order the inverse of the size of the system) using Bethe Ansatz. Starting from the functional formulation of the Bethe equations, we obtain for all the cumulants of the current both the leading and next-to-leading contribution in the size of the system.
\end{abstract}

\maketitle

\begin{section}{Introduction}
The one dimensional asymmetric simple exclusion process (ASEP) is one of the most simple examples of a classical interacting particles system exhibiting a non equilibrium steady state. This stochastic system has been studied much in the past, both in the mathematical \cite{S70.1,L85.1,F08.1} and physical literature \cite{KLS84.1,S91.1,HHZ95.1,SZ95.1,D98.1,S01.1,GM06.1,D07.1}. It consists of particles hopping locally on a one dimensional lattice, with an asymmetry between the forward and backward hopping rates. The exclusion constraint prevents the particles from moving to a site already occupied by another particle. The asymmetry between the hopping rates models the action of an external driving field in the bulk of the system, which maintains a permanent macroscopic current in the system. This current breaks the detailed balance and keeps the system out of equilibrium even in the stationary state. The special case for which the particles hop forward and backward with equal rates is called the symmetric simple exclusion process (SSEP). It corresponds to a situation for which the detailed balance holds in the bulk which means (in the absence of boundary conditions breaking the forward-backward symmetry) that the system reaches equilibrium in the long time limit. In this case, the system belongs to the universality class of the Edwards--Wilkinson (EW) equation \cite{EW82.1,HHZ95.1}. On the contrary, if the two hopping rates are different, detailed balance is broken and the system reaches in the long time limit a non equilibrium steady state characterized by the presence of a current of particles flowing through the system. In that case, the system belongs to the universality class of the Kardar--Parisi--Zhang (KPZ) equation \cite{KPZ86.1,HHZ95.1}.\\\indent
Because of its simplicity, the ASEP is an interesting tool to investigate the general properties of systems out of equilibrium. Moreover, the ASEP is related through various mappings to many other models, in particular: the zero range process \cite{EH05.1}, directed polymers in a random medium \cite{K97.1}, interface growth models \cite{GS92.1,HHZ95.1,K97.1}, the six vertex model \cite{KDN90.1,GS92.1}, XXZ spin chains \cite{GS92.1,ER96.1,GM06.1}. It is also used as a starting point to model physical phenomena such as cellular molecular motors \cite{LKN01.1}, hopping conductivity \cite{R77.1}, traffic flow \cite{CSS00.1}, usually by enriching the dynamics of the ASEP by new rules that makes it closer to the studied phenomenon.\\\indent
The ASEP, along with a very small number of other statistical mechanics models is known to be ``exactly solvable'' in the sense that several quantities can be calculated exactly, which is a rather uncommon property. The totally asymmetric case, for which all the particles hop in only one direction (TASEP), is usually the easiest to solve. On the contrary, the model with partial asymmetry often exhibits a more intricate mathematical structure and is thus more difficult to solve. A few different approaches have been used in the past to obtain exact results for the ASEP: the matrix product representation \cite{DEHP93.1,BE07.1} allows to calculate explicitly the probabilities of all the configurations in the stationary state for both open systems connected to reservoirs of particles and systems on a ring with periodic boundary conditions; techniques from random matrix theory \cite{PS02.1,S06.1,S07.1} have been used for the study of infinite systems defined on $\mathbb{Z}$; Bethe Ansatz has provided many exact results, principally on a ring \cite{D87.1,GS92.1}, but also recently for open systems \cite{dGE05.1}.\\\indent
The fact that Bethe Ansatz can be used to study the ASEP is strongly related to the ``quantum integrability'' of the model. Indeed, the Markov matrix governing the time evolution of the probabilities for the configurations of the ASEP is similar to the hamiltonian of the XXZ spin chain and closely related to the transfer matrix of the six vertex model. While its formulation for the ASEP is well understood, the use of the Bethe Ansatz is usually quite technical. The difficulty lies in the determination of the so called ``Bethe roots'' in terms of which all the quantities we want to calculate are expressed. These Bethe roots are solutions of a set of highly coupled polynomial equations called the Bethe equations of the system. Their solutions are usually not known for general values of the parameters of the model studied. In the case of the ASEP, Bethe Ansatz has been used successfully to calculate the gap of the system \cite{GS92.1,K95.1,GM04.1,dGE05.1,dGE08.1}, related to the dynamical exponent which governs the speed at which the system reaches its stationary state. It has also allowed to calculate some properties of the fluctuations of the current in the stationary state \cite{DL98.1,DA99.1,LK99.1,DE99.1,ADLvW08.1,PM08.1,P08.1}.\\\indent
In the present work, we study the fluctuations of the steady state current for the ASEP on a ring in the scaling of a weak asymmetry between the hopping rates (asymmetry scaling as the inverse of the size of the system). The main result of the article is the Bethe Ansatz derivation of all the cumulants of the current in this scaling limit. We obtain an explicit expression (\ref{Etilde(mu,nu)}) for the leading and next-to-leading contributions in the size of the system. We use a rewriting of the Bethe equations for the ASEP in terms of a polynomial functional equation. Our approach is based on the functional Bethe Ansatz and does not rely on the behavior of the Bethe roots in the large system size limit. We check our results numerically by solving the functional Bethe equation for systems up to size $100$.\\\indent
In section \ref{section discussion WASEP}, we discuss our formula (\ref{Etilde(mu,nu)}) for the cumulants of the current in the scaling of a weak asymmetry. In section \ref{section ebQR}, we write the Bethe equations for the ASEP as a polynomial functional equation and recall how this equation can be solved perturbatively to calculate the first cumulants of the current. In section \ref{section regularisation}, we define a new version of the functional equation that remains regular in the symmetric limit. Then, in section \ref{section limite WASEP}, we take the weakly asymmetric limit of this equation, and we solve it in section \ref{section solution WASEP}. A few technical calculations are relegated to the appendix.
\end{section}

\begin{section}{Cumulants of the current in the weakly asymmetric exclusion process}
\label{section discussion WASEP}
We consider the partially asymmetric simple exclusion process (PASEP) on a ring. It is a stochastic process involving classical hard core particles hopping on a one dimensional lattice with periodic boundary conditions. Each one of the $L$ sites can be occupied by at most one of the $n$ particles. The system evolves with the following local dynamics: in an infinitesimal time interval $dt$, each particle hops forward with probability $p\,dt$ and backward with probability $p\,x\,dt$ if the destination site is empty (exclusion rule).

\begin{subsection}{Fluctuations of the total current}
We define the total integrated current $Y_{t}$ between time $0$ and time $t$ as the total distance covered by all the particles in this duration. This is a random variable which depends on the trajectories of the particles starting in some configuration $\mathcal{C}$ at time $0$ and evolving by the markovian dynamics up to time $t$. We are interested in the fluctuations of $Y_{t}$ in the long time limit, when the (finite size) system reaches its unique stationary state which is independent of the initial configuration $\mathcal{C}$. We emphasize that we consider here the long time limit for finite systems. We will take the large system size limit only in the end. This is a completely different regime from taking the infinite system size limit $L\to\infty$ first and studying then the stationary state $t\to\infty$ \cite{HHZ95.1}. We want to calculate the first cumulants of the current with respect to the stationary state probability distribution of $Y_{t}$, that is its mean value $J(x)$, the diffusion constant $D(x)$ and the higher cumulants:
\begin{align}
&J(x)=\lim_{t\to\infty}\frac{\langle Y_{t}\rangle}{t}\\
&D(x)=\lim_{t\to\infty}\frac{\langle Y_{t}\rangle^{2}-\langle Y_{t}^{2}\rangle}{t}\\
&E_{3}(x)=\lim_{t\to\infty}\frac{\langle Y_{t}^{3}\rangle-3\langle Y_{t}\rangle\langle Y_{t}^{2}\rangle+2\langle Y_{t}\rangle^{3}}{t}\;.
\end{align}
The characteristic function, defined as the mean value of $e^{\gamma Y_{t}}$ behaves in the long time limit as \cite{PM08.1}
\begin{equation}
\left\langle e^{\gamma Y_{t}}\right\rangle\sim e^{E(\gamma,x)t}\;,
\end{equation}
where $\gamma$ is the ``fugacity'' corresponding to the variable $Y_{t}$. Taking the derivatives of the previous expression with respect to $\gamma$, we see that $E(\gamma,x)$ is the exponential generating function of the cumulants of the current in the stationary state, \textit{i.e.}
\begin{equation}
\label{E(J,D,E3)}
E(\gamma,x)=J(x)\gamma+\frac{D(x)}{2!}\,\gamma^{2}+\frac{E_{3}(x)}{3!}\,\gamma^{3}+...
\end{equation}
The generating function of the cumulants $E(\gamma,x)$ can be related to the large deviation function of the current $G(j,x)$. The function $G(j,x)$ is defined from the asymptotic behavior of the probability distribution of $Y_{t}$ as
\begin{equation}
P_{t}(j)\equiv P\left(\frac{Y_{t}}{t}=j\right)\sim e^{-G(j,x)t}\quad\text{when $t\to\infty$}\;.
\end{equation}
The large deviation function of the current $G(j,x)$ is equal to $0$ for $j=J(x)$ and is strictly positive otherwise, leading to an exponentially vanishing probability in the long time limit except when $j$ is equal to the mean value of the current. Writing
\begin{equation}
\left\langle e^{\gamma Y_{t}}\right\rangle=\int dj\,e^{j\gamma t}P_{t}(j)\sim\int dj e^{(j\gamma-G(j,x))t}\sim e^{t\max_{j}(j\gamma-G(j,x))}\;,
\end{equation}
we observe that $E(\gamma,x)$ is the Legendre transform of the large deviations function $G(j,x)$, that is
\begin{equation}
E(\gamma,x)=\max_{j}(j\gamma-G(j,x))\;.
\end{equation}
\end{subsection}

\begin{subsection}{Cumulants of the current in the scaling of a weak asymmetry}
The principal result of this article is the calculation of all the cumulants of the stationary state current in the weakly asymmetric scaling limit $1-x\sim 1/L$, or equivalently, the Taylor expansion in the vicinity of $\gamma=0$ of the generating function $E(\gamma,x)$. Using Bethe Ansatz, we find
\begin{align}
\label{Etilde(mu,nu)}
\frac{1}{p}\,\tilde{E}(\mu,\nu)&\equiv\frac{1}{p}\,E\left(\gamma=\frac{\mu}{L},x=1-\frac{\nu}{L}\right)\\
&=\frac{\rho(1-\rho)(\mu^{2}+\mu\nu)}{L}+\frac{1}{L^{2}}\left(-\frac{\rho(1-\rho)\mu^{2}\nu}{2}+\varphi[\rho(1-\rho)(\mu^{2}+\mu\nu)]\right)+\O{\frac{1}{L^{3}}}\;.\nonumber
\end{align}
In this expression, $\rho$ is the particle density $\rho=n/L$ and the function $\varphi$ is given by
\begin{equation}
\varphi(z)=\sum_{k=1}^{\infty}\frac{B_{2k-2}}{k!(k-1)!}z^{k}\;,
\end{equation}
where the $B_{j}$ are the Bernoulli numbers. The expression (\ref{Etilde(mu,nu)}) gives the leading and next-to-leading order in $L$ of all the cumulants of the current by taking the derivative with respect to $\mu$. Only the subleading term of (\ref{Etilde(mu,nu)}) contributes to the $k$-th cumulant $E_{k}$ for $k\geq 3$. In this case, $E_{k}$ is a polynomial of degree $k$ in the rescaled asymmetry $\nu$. The coefficients of this polynomial are expressed in terms of the Bernoulli numbers multiplied by factorials. The first cumulants $E_{k}$ are plotted with respect to the rescaled asymmetry $\nu$ in fig. \ref{fig oscillations cumulants}. We note that they show oscillations in the parameter $\nu$. This oscillation phenomenon of the cumulants of the current has been observed recently in the context of electron transport through a quantum dot in \cite{FFHNNBH09.1}. The special case $\nu=0$ which corresponds to the SSEP has already been calculated by Bethe Ansatz in \cite{ADLvW08.1}. For arbitrary $\nu$, equation (\ref{Etilde(mu,nu)}) leads to an expression for the large deviation function $G$ up to the order $2$ in $L$, which matches the result obtained in \cite{ADLvW08.1} using the macroscopic fluctuation theory developed in \cite{BDSGJLL01.1,BDSGJLL04.1}.\\\indent
\begin{figure}
\begin{tabular}{ccccc}
\!\!\!\!\!\!\!\!\!\!\!\!\includegraphics{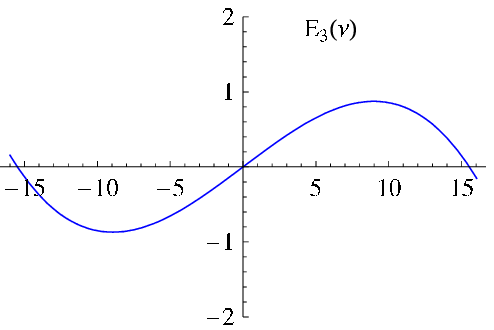} & & \includegraphics{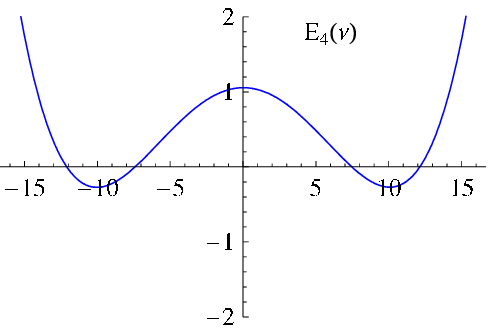} & & \includegraphics{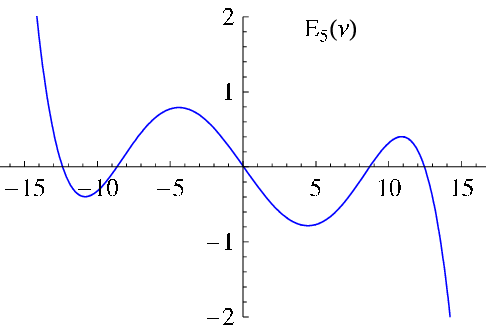}\\
 & & & & \\
\!\!\!\!\!\!\!\!\!\!\!\!\includegraphics{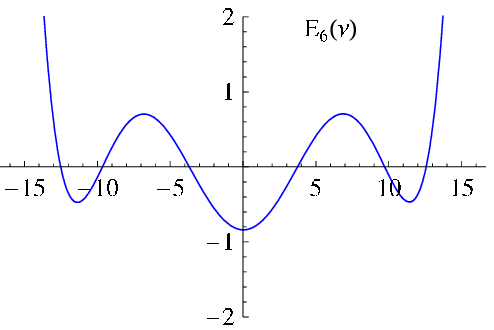} & & \includegraphics{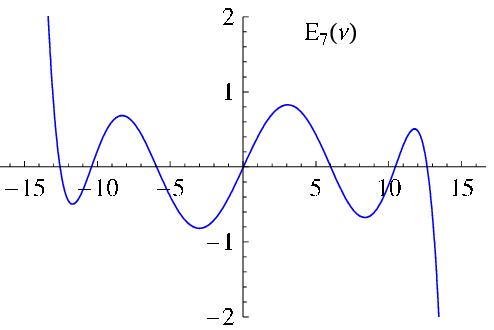} & & \includegraphics{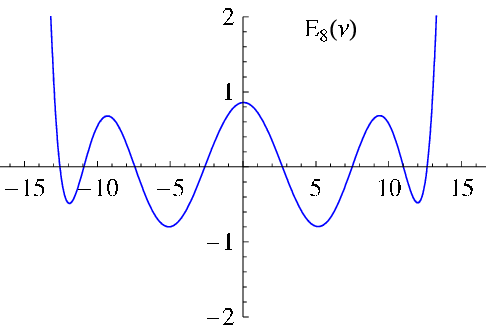}\\
 & & & & \\
\!\!\!\!\!\!\!\!\!\!\!\!\includegraphics{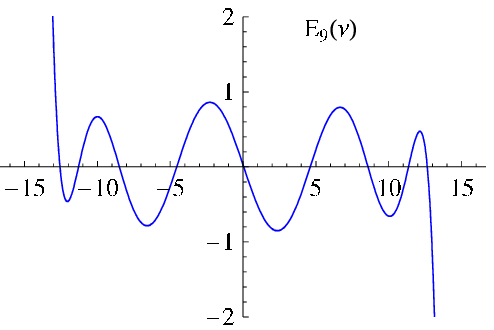} & & \includegraphics{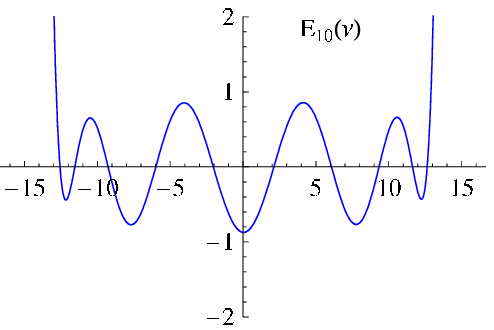} & & \includegraphics{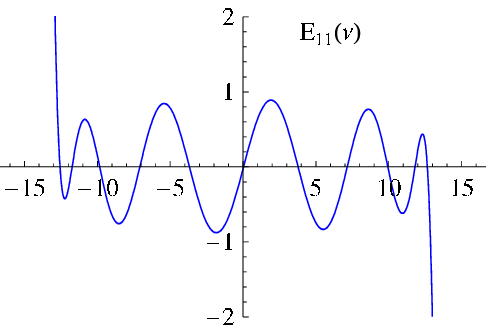}\\
\end{tabular}
\caption{Graphs of the first cumulants of the current $E_{k}$ in the weakly asymmetric scaling limit, obtained by taking the successive derivatives in $\mu$ of the generating function (\ref{Etilde(mu,nu)}) at $\mu=0$. The rescaled cumulants $\frac{k-2}{8\pi^{2}}(\pi e/(kL\sqrt{\rho(1-\rho)})^{k}E_{k}(\nu)$ are plotted with respect to the rescaled asymmetry $\nu$ at particle density $\rho=1/2$ for $k$ from $3$ to $11$.}
\label{fig oscillations cumulants}
\end{figure}
We now justify the weakly asymmetric scaling chosen in equation (\ref{Etilde(mu,nu)}). The crossover between the Edwards--Wilkinson and the Kardar--Parisi--Zhang behavior lies at a scaling where the asymmetry $1-x$ is nonzero but goes to zero when the size $L$ of the system goes to infinity. Consider a tagged particle in the system with asymmetry scaling as $1-x\sim 1/L^{r}$. During a time interval $\Delta t$, this particle makes a number of rotations $R\sim(1-x)\Delta t/L$ through the system. A typical time interval $\Delta t$ to consider is the time necessary for the system to reach its stationary state: $\Delta t\sim L^{z}$, $z$ being the dynamic exponent of the system. Then, a criterion for separation between weak and strong asymmetry is when $R\sim 1$: when $R\gg 1$, the system is asymmetric, while $R\ll 1$ corresponds to a symmetric system. This leads to $1-x\sim 1/L^{z-1}$. The value of $z$ depends on whether the system is in the Edwards--Wilkinson (EW, $z=2$) or in the Kardar--Parisi--Zhang (KPZ, $z=3/2$) universality class. This leads to two natural scalings for the asymmetry, $1-x\sim 1/L$ and $1-x\sim 1/\sqrt{L}$. It turns out that both of these scalings are meaningful for the ASEP. The scaling $1-x\sim 1/L$ corresponds to the weakly asymmetric exclusion process, whereas the scaling $1-x\sim1/\sqrt{L}$ corresponds to the crossover between the EW and the KPZ regimes \cite{K95.1}. In particular, the weakly asymmetric scaling $1-x\sim 1/L$ belongs to the EW class like the symmetric exclusion process.\\\indent
From the formula (\ref{Etilde(mu,nu)}), we observe that $\tilde{E}(\mu,\nu)$ for $\nu\neq 0$ is a rather minimal deformation of the case $\nu=0$. This modification is in fact needed to ensure that $\tilde{E}(\mu,\nu)$ stays invariant under the Gallavotti--Cohen symmetry given by \cite{LS99.1,GM06.1}
\begin{equation}
E(\gamma,x)=E(\log x-\gamma,x)\;,
\end{equation}
which in the weakly asymmetric scaling leads to
\begin{equation}
\tilde{E}(\mu,\nu)=\tilde{E}\left(-\mu-\nu-\frac{\nu^{2}}{2L}+\O{\frac{1}{L^{2}}},\nu\right)\;
\end{equation}
for the leading and next-to-leading order. The study of the exact values for the diffusion constant \cite{DM97.1,PM08.1} and the third cumulant \cite{P08.1} shows that these two cumulants are still given by equation (\ref{Etilde(mu,nu)}) as long as $1-x\ll 1/\sqrt{L}$. This suggests that this minimal deformation is valid in all the Edwards--Wilkinson universality class.
\end{subsection}

\begin{subsection}{Phase transition}
In \cite{BD05.1}, using the macroscopic fluctuation theory for driven diffusive systems developed in \cite{BDSGJLL01.1,BDSGJLL04.1}, the existence of a nontrivial dynamical phase transition was found in the weakly asymmetric exclusion process, with in particular a phase of weaker asymmetry for which the fluctuations of the current are gaussian (at dominant order in $1/L$), and a phase of stronger asymmetry in which the fluctuations of the current become non gaussian. Let $\nu_{c}$ be the value of the rescaled asymmetry corresponding to the separation between the gaussian and non gaussian phases. For $\nu<\nu_{c}$, the large deviation function of the current $G(j,\nu)$ (respectively the generating function of the cumulants of the current $\tilde{E}(\mu,\nu)$) is quadratic in $j$ (resp. $\mu$) at the leading order in the size of the system. Performing the Legendre transform of the leading order of the expression (\ref{Etilde(mu,nu)}) for $\tilde{E}(\mu,\nu)$, we find that in the gaussian phase $\nu<\nu_{c}$, the large deviation function $G(j,\nu)$ is given by the quadratic function of $j$:
\begin{equation}
\label{G(j) quadratic}
G(j,\nu)=\frac{(j-J)^{2}}{2D}\;,
\end{equation}
with
\begin{align}
\frac{J}{p}&=\rho(1-\rho)\nu\\
\frac{D}{p}&=2\rho(1-\rho)L
\end{align}
at the leading order in $L$. On the contrary, in the non gaussian phase $\nu>\nu_{c}$, neither $G(j,\nu)$ nor $\tilde{E}(\mu,\nu)$ are expected to be quadratic, even at the leading order in $L$. We emphasize that it does not contradict the fact that the Taylor expansion of $\tilde{E}(\mu,\nu)$ given in equation (\ref{Etilde(mu,nu)}) is quadratic in $\mu$ at the leading order. It merely means that the large $L$ limit of the asymptotic formula for $\tilde{E}(\mu,\nu)$ breaks down and does not represent the full function $\tilde{E}(\mu,\nu)$ anymore. We will come back to this issue at the end of this subsection.\\\indent
According to \cite{BD05.1}, the density profile adopted by the system is dependent on the value of the current flowing through the system. In the gaussian phase, the density profile remains flat for all values of the current. In the non gaussian phase however, the density profile depends on the value of the current: there is a critical value $j_{c}(\nu)$ for the current such that if $|j|>j_{c}(\nu)$, the density profile remains flat, while if $|j|<j_{c}(\nu)$, the profile becomes time dependent. The signature of this transition between the flat profile and the time dependent profile is visible through the appearance of non analyticities in the large deviation function of the current or, equivalently, in its Legendre transform, the (rescaled) generating function of the cumulants $\tilde{E}(\mu,\nu)$. A non analyticity in the large deviation function $G(j,\nu)$ at $j=\pm j_{c}(\nu)$ corresponds to a non analyticity in $\tilde{E}(\mu,\nu)$ at $\mu_{c,1}(\nu)$ and $\mu_{c,2}(\nu)$ related through the Gallavotti--Cohen symmetry as $\mu_{c,1}(\nu)+\mu_{c,2}(\nu)=L\log(1-\nu/L)\sim -\nu$.\\\indent
We now look at the non analyticities of the expression (\ref{Etilde(mu,nu)}) for $\tilde{E}(\mu,\nu)$. From the asymptotic behavior of the Bernoulli numbers
\begin{equation}
B_{2k}\sim(-1)^{k-1}4\sqrt{\pi k}\left(\frac{k}{\pi e}\right)^{2k}\;,
\end{equation}
we observe that $\varphi(z)$ has a singularity at $z=-\pi^{2}$. It corresponds for the function $\tilde{E}(\mu,\nu)$ to singularities at the points $\rho(1-\rho)(\mu^{2}+\mu\nu)=-\pi^{2}$. This equation has real solutions for $\mu$ if $\rho(1-\rho)\nu^{2}>4\pi^{2}$. Thus, non analyticities appears in $\tilde{E}(\mu,\nu)$ as soon as $\nu>\nu_{c}$ for
\begin{equation}
\nu_{c}=\frac{2\pi}{\sqrt{\rho(1-\rho)}}\;,
\end{equation}
and in this case, the non analyticities of $\tilde{E}(\mu,\nu)$ are at the points
\begin{equation}
\mu_{c,1|2}(\nu)=\frac{-\nu\pm\sqrt{\nu^{2}-\frac{4\pi^{2}}{\rho(1-\rho)}}}{2}\;.
\end{equation}
These expressions for $\mu_{c,1}(\nu)$ and $\mu_{c,2}(\nu)$ should hold only in the vicinity of $\nu=\nu_{c}$ since we used the expression (\ref{Etilde(mu,nu)}) for $\tilde{E}(\mu,\nu)$ which is valid for $\mu$ far from $0$ only in the gaussian phase. For the Legendre transform of the gaussian leading order of (\ref{Etilde(mu,nu)}), we define the function $\mu(j)$ such that
\begin{equation}
j=L\frac{d}{d\mu}\tilde{E}(\mu(j),\nu)\quad\text{and}\quad \tilde{E}(\mu(j),\nu)+G(j,\nu)=j\,\frac{\mu(j)}{L}
\end{equation}
By the inverse function of $\mu(j)$, the values $\mu_{c,1}(\nu)$ and $\mu_{c,2}(\nu)$ correspond for the large deviation function $G(j,\nu)$ to $\pm j_{c}(\nu)$ with
\begin{equation}
\frac{j_{c}(\nu)}{p}=\rho(1-\rho)\sqrt{\nu^{2}-\frac{4\pi^{2}}{\rho(1-\rho)}}
\end{equation}
near $\nu=\nu_{c}$. By Legendre transform, the function $\mu(j)$ sends the region $|j|<j_{c}(\nu)$ (where the density profile is time dependent) of the plane $(j,\nu)$ to the region $\mu_{c,1}(\nu)<\mu<\mu_{c,2}(\nu)$, or equivalently $\rho(1-\rho)(\mu^{2}+\mu\nu)<-\pi^{2}$, of the plane $(\mu,\nu)$. It gives at the leading order of (\ref{Etilde(mu,nu)})
\begin{equation}
L\frac{\tilde{E}(\mu,\nu)}{p}<-\pi^{2}\;,
\end{equation}
which agrees with equation (25) of \cite{BD05.1} (where $p$ is taken to be equal to $1/2$).\\\indent
\begin{figure}
\begin{tabular}{ccc}
\!\!\!\!\!\!\!\!\!\!\!\!\includegraphics{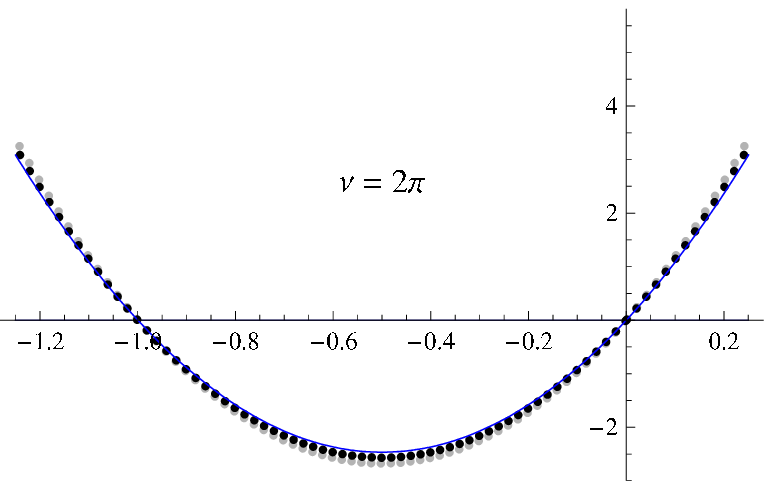} & & \includegraphics{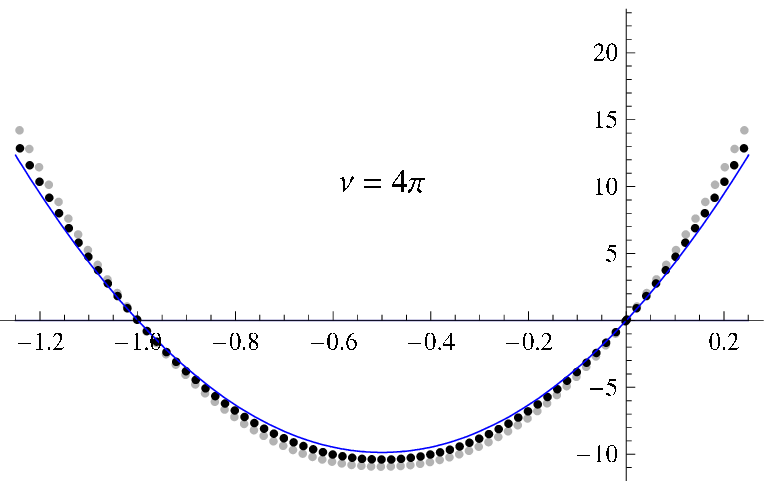}\\
 & & \\
\!\!\!\!\!\!\!\!\!\!\!\!\includegraphics{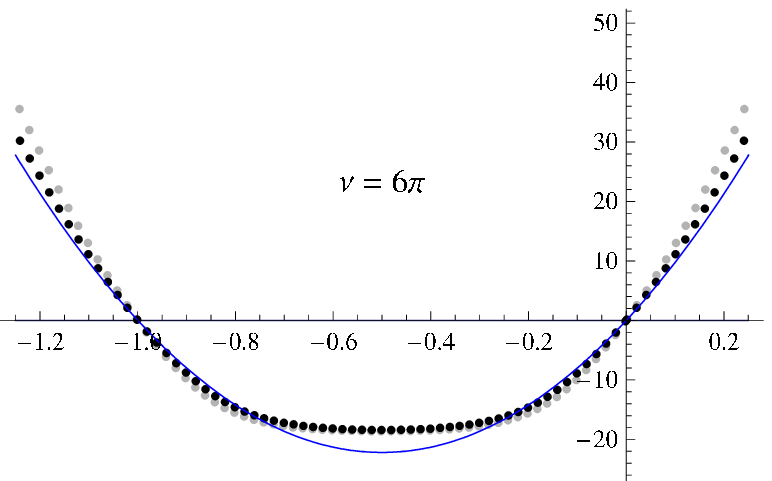} & & \includegraphics{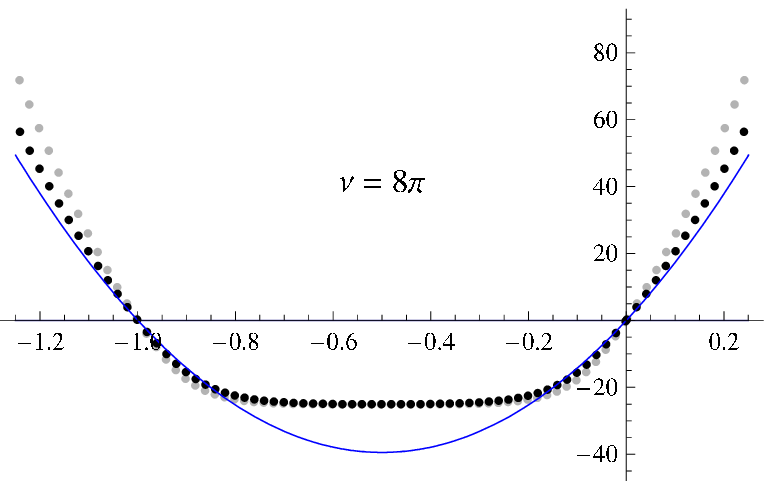}\\
 & & \\
\!\!\!\!\!\!\!\!\!\!\!\!\includegraphics{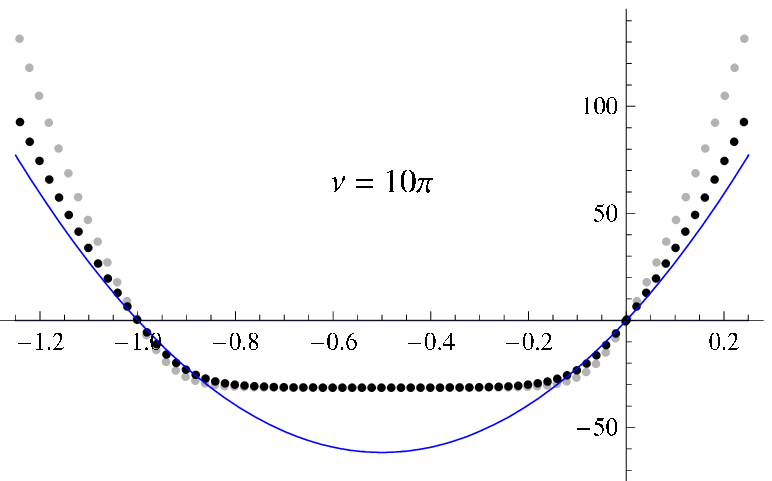} & &
\end{tabular}
\caption{Rescaled generating function of the cumulants of the current $L\tilde{E}(\mu,\nu)$ at half filling, plotted with respect to $w=\mu/(L\log{x})$ between $w=-1.25$ and its symmetric through the Gallavotti--Cohen symmetry $w=0.25$ for $\nu=2\pi$, $\nu=4\pi$, $\nu=6\pi$, $\nu=8\pi$ and $\nu=10\pi$. The solid line represents the leading order of the result (\ref{Etilde(mu,nu)}) of the Bethe Ansatz calculation. The gray dots correspond to the numerical resolution of the functional Bethe equation for $n=25$, $L=50$. The black dots correspond to the numerical resolution of the functional Bethe equation for $n=50$, $L=100$.}
\label{fig E(gamma,x) numerical}
\end{figure}
Our resolution of the functional Bethe equation of the ASEP (\ref{ebQR}) in the weakly asymmetric scaling limit, leading to the expression (\ref{Etilde(mu,nu)}) for the cumulants of the current, relies on a perturbative expansion near $\mu=0$ of the functional Bethe equation. This approach does not always give an information on the value of $\tilde{E}(\mu,\nu)$ for $\mu$ far from $0$.\\\indent
From the discussion in the beginning of this subsection of the gaussian/non-gaussian phase transition, it is expected that the function $\tilde{E}(\mu,\nu)$ will be equal to its Taylor expansion for $\nu<\nu_{c}$. In particular, the function should be quadratic in $\mu$ in the large $L$ limit. On the contrary, for $\nu>\nu_{c}$, $\tilde{E}(\mu,\nu)$ is expected to be different from its Taylor expansion (\ref{Etilde(mu,nu)}), even at the leading order in $L$.\\\indent
In order to check whether the function $\tilde{E}(\mu,\nu)$ was equal to its Taylor expansion in $\mu=0$ given by (\ref{Etilde(mu,nu)}), we studied numerically the Bethe equations of the ASEP for systems up to size $L=100$ (see appendix \ref{appendix numerical}). In fig. \ref{fig E(gamma,x) numerical}, we show the results we obtained at half filling ($\nu_{c}=4\pi$) for different values of the asymmetry $\nu$ and of the size of the system $L$. These results are in excellent agreement with the emergence of non gaussianity for $\nu>4\pi$: for $0<\nu<4\pi$, the numerical evaluation of $E(\gamma,x)$ fits well with the quadratic expression given by the leading order of equation (\ref{Etilde(mu,nu)}). For $\nu>4\pi$, however, there is a region between $\gamma=0$ and its symmetric by the Gallavotti--Cohen symmetry $\gamma=\log x$ where $E(\gamma,x)$ differs significantly from the quadratic leading order of (\ref{Etilde(mu,nu)}). Outside of this region, the numerical evaluation of $E(\gamma,x)$ still agrees with (\ref{Etilde(mu,nu)}).\\\indent
We emphasize that the fact $\tilde{E}(\mu,\nu)$ is defined for a finite system in (\ref{E(J,D,E3)}) and (\ref{Etilde(mu,nu)}) as a generating function in $\mu$, that is, as a Taylor series for $\mu$ at $\mu=0$, does not contradict the fact that $\tilde{E}(\mu,\nu)$ can be different from its Taylor expansion at $\mu=0$ in the large $L$ limit. An example of such a behavior is exhibited by the function $\sum_{j=0}^{L}(-L^{2}/(1+L^{2}\mu^{2}))^{j}/j!=e^{-1/\mu^{2}}+\O{1/L}$. This function of $\mu$ and $L$ is, for finite $L$, a rational fraction in $\mu$ which is entirely defined for $\mu\in\mathbb{C}$ by its Taylor expansion in $\mu=0$ through an analytic continuation, but develops an essential singularity in $\mu=0$ when $L\to\infty$.
\end{subsection}
\end{section}

\begin{section}{Reminder of the functional formulation of the Bethe equations}
\label{section ebQR}
In this section, we recall the functional formulation of the Bethe Ansatz for the ASEP. We show how the generating function of the cumulants (\ref{E(J,D,E3)}) can be expressed in terms of a solution to a functional polynomial equation and how this functional equation can be solved perturbatively to obtain the first cumulants.\\\indent
It can be shown \cite{DL98.1,PM08.1} that the generating function $E(\gamma,x)$ of the cumulants of $Y_{t}$ over the variable $\gamma$ is equal to the eigenvalue with largest real part of a deformation $M(\gamma)$ of the Markov matrix $M$ of the system. Because of the underlying integrability of the ASEP, the diagonalization of $M(\gamma)$ can be performed using the Bethe Ansatz. The Bethe equations of the system can be rewritten \cite{PM08.1} in the functional equation
\begin{equation}
\label{ebQR}
Q(t)R(t)=e^{L\gamma}(1-t)^{L}Q(xt)+(1-xt)^{L}x^{n}Q(t/x)\;.
\end{equation}
The polynomial $Q$ is of degree $n$ and the polynomial $R$ is of degree $L$. We choose the normalization of $Q$ such that the coefficient of highest degree of $Q(t)$ is equal to one:
\begin{equation}
Q(t)=\prod_{j=1}^{n}(t-y_{j})\;.
\end{equation}
If we set $t=y_{i}$ in the functional Bethe equation (\ref{ebQR}), we obtain
\begin{equation}
\label{eby}
e^{L\gamma}\left(\frac{1-y_{i}}{1-xy_{i}}\right)^{L}=-\prod_{j=1}^{n}\frac{y_{i}-xy_{j}}{xy_{i}-y_{j}}\;,
\end{equation}
which is the usual form of the Bethe equations in terms of the Bethe roots $y_{i}$. Equations (\ref{ebQR}) and (\ref{eby}) are completely equivalent forms of the Bethe equations. In particular, they both have a large discrete set of solutions corresponding to the different eigenstates of the deformed matrix $M(\gamma)$. We are only interested in the solution corresponding to the largest eigenvalue of $M(\gamma)$, which is characterized by
\begin{equation}
\label{Q(gamma=0)}
Q(t)=t^{n}+\O{\gamma}\;.
\end{equation}
Equivalently, the Bethe roots $y_{i}$ all tend to $0$ when $\gamma\to 0$ for this solution of the Bethe equations. In the following, we will also use a relation coming from the fact that the stationary state is a zero momentum state \cite{PM08.1}. This condition implies
\begin{equation}
\label{Q(1)}
e^{n\gamma}Q(1)=x^{n}Q(1/x)\;.
\end{equation}
The generating function $E(\gamma,x)$ of the cumulants of the current is then given by \cite{PM08.1}
\begin{equation}
\label{E[Q]}
\frac{E(\gamma,x)}{p}=(1-x)\left(\frac{Q'(1)}{Q(1)}-\frac{1}{x}\,\frac{Q'(1/x)}{Q(1/x)}\right)=(1-x)\sum_{i=1}^{n}\left(\frac{1}{1-y_{i}}-\frac{1}{1-xy_{i}}\right)\;.
\end{equation}
In the rest of this section, we explain how the functional Bethe equation can be solved perturbatively near $\gamma=0$. Introducing the function
\begin{equation}
\label{A[Q]}
A(t)=\frac{x^{n}Q(t/x)}{e^{n\gamma}Q(t)}\;,
\end{equation}
the functional Bethe equation (\ref{ebQR}) becomes
\begin{equation}
\label{R[A]}
e^{-n\gamma}R(t)=(1-xt)^{L}A(t)+(1-t)^{L}\frac{x^{n}e^{(L-2n)\gamma}}{A(xt)}\;.
\end{equation}
Note that we have added an extra factor $e^{n\gamma}$ in the definition compared to \cite{P08.1}. In terms of $A(t)$, equations (\ref{Q(gamma=0)}), (\ref{Q(1)}) and (\ref{E[Q]}) become respectively
\begin{align}
\label{A(gamma=0)}
A(t)&=1+\O{\gamma}\\
\label{A(1)}
A(1)&=1\\
\label{E[A]}
\frac{E(\gamma,x)}{p}&=-(1-x)A'(1)\;.
\end{align}
In the following, we will also need a few other properties of $A(t)$. Because $Q$ is a polynomial of degree $n$, we infer from the definition (\ref{A[Q]}) of $A(t)$ that
\begin{equation}
\label{A(infinity)}
\lim_{t\to\infty}A(t)=e^{-n\gamma}\;.
\end{equation}
This equation fixes the term of highest degree of the polynomial $R$, using the relation (\ref{R[A]}) between $R(t)$ and $A(t)$:
\begin{equation}
\label{R(infinity)}
e^{-n\gamma}R(t)-(x^{L}e^{-n\gamma}+x^{n}e^{(L-n)\gamma})(-1)^{L}t^{L}\qquad\text{is a polynomial in $t$ of degree $L-1$.}
\end{equation}
Since the value of $A(t)$ is known when $\gamma$ is equal to zero (\ref{A(gamma=0)}), it is natural to attempt solving the functional Bethe equation (\ref{R[A]}) perturbatively near $\gamma=0$. Moreover, a perturbative solution of (\ref{R[A]}) in powers of $\gamma$ gives access to the first cumulants of the current. Using (\ref{A(gamma=0)}), we write the expansion of $A(t)$ in powers of $\gamma$ as
\begin{equation}
\label{Ak}
A(t)=1+\sum_{k=1}^{\infty}A_{k}(t)\gamma^{k}\;.
\end{equation}
From (\ref{Q(gamma=0)}) and the definition (\ref{A[Q]}) of $A(t)$, we deduce that the $A_{k}(t)$ are polynomials in $1/t$ of degree $kn$. This observation will be crucial for the following. We will now see that equation (\ref{R[A]}) can be reformulated as a recurrence formula which can be used to calculate explicitly the first $A_{k}(t)$ and, as a consequence, the first cumulants. In particular, the three first cumulants were calculated in \cite{P08.1} for finite size systems by this method that we now recall (see \cite{P08.1} for more details). Reminding that $R(t)$ is a polynomial in $t$, that is $R(t)$ has only nonnegative powers in $t$, we can eliminate it from equation (\ref{R[A]}) by doing the expansion for $t\to 0$. We have
\begin{equation}
\label{recurrA1}
\frac{A(t)}{(1-t)^{L}}+\frac{1}{(1-xt)^{L}}\frac{x^{n}e^{(L-2n)\gamma}}{A(xt)}=\O{t^{0}}\;.
\end{equation}
This equation must be understood in the following way: first, we expand the l.h.s. around $\gamma=0$ in terms of the $A_{k}(t)$. Then, at each order in $\gamma$, the l.h.s. has a finite limit when $t\to 0$, that is all the negative powers in $t$ from the $A_{k}(t)$ cancel out. It is crucial to respect the order between the two expansions in $\gamma$ and $t$. Expanding first around $\gamma=0$ makes the poles of $A(t)$ (that is, the Bethe roots $y_{i}$) disappear from the problem, leaving us only with the algebraic properties of the polynomials $A_{k}(t)$. On the contrary, if we did the expansion in $t=0$ first, equation (\ref{recurrA1}) would not contain any information since $A(t)$ is regular when $t\to 0$. Introducing the operator $\Delta_{x}$ which acts on an arbitrary function $u$ as
\begin{equation}
\left(\Delta_{x}u\right)(t)=u(t)-x^{n}u(xt)\;,
\end{equation}
we rewrite the previous equation in the slightly more complicated form
\begin{equation}
\label{recurrA2}
\Delta_{x}\left(\frac{A(t)}{(1-t)^{L}}\right)=-\frac{x^{n}}{(1-xt)^{L}}\left(A(xt)+\frac{e^{(L-2n)\gamma}}{A(xt)}\right)+\O{t^{0}}\;.
\end{equation}
At order $k$ in $\gamma$, the r.h.s. depends only on the $A_{j}(t)$ for $j<k$. We emphasize that $A_{k}(t)$ cancels out. This observation is the key that allows us to solve $A(t)$ order by order in $\gamma$. Noting that $\Delta_{x}$ acts separately on each power of $t$ and that $\Delta_{x}\O{t^{0}}=\O{t^{0}}$, we can invert $\Delta_{x}$ in (\ref{recurrA2}). We have
\begin{equation}
\label{recurrA3}
\frac{A(t)}{(1-t)^{L}}=-\Delta_{x}^{-1}\left(\frac{x^{n}}{(1-xt)^{L}}\left(A(xt)+\frac{e^{(L-2n)\gamma}}{A(xt)}\right)\right)-b\,\frac{1}{t^{n}}+\O{t^{0}}\;.
\end{equation}
The additional term $b/t^{n}$ comes from the fact that the operator $\Delta_{x}$ gives $0$ when applied on $1/t^{n}$. Recalling (\ref{A(infinity)}) and the fact that the $A_{k}(t)$ have only negative (or zero) powers in $t$, we finally obtain
\begin{equation}
\label{recurrA4}
A(t)=e^{-n\gamma}-\left[(1-t)^{L}\Delta_{x}^{-1}\left(\frac{x^{n}}{(1-xt)^{L}}\left(A(xt)+\frac{e^{(L-2n)\gamma}}{A(xt)}\right)\right)\right]_{(t^{-\infty})}^{(t^{-1})}-b\left[\frac{(1-t)^{L}}{t^{n}}\right]_{(t^{-\infty})}^{(t^{-1})}\;.
\end{equation}
We used the notation $[u(t)]_{(t^{-\infty})}^{(t^{-1})}$ for the negative powers in $t$ of a function $u(t)$ in its expansion near $t=0$, after the perturbative expansion near $\gamma=0$ as before. The constant $b$ can be set, order by order in $\gamma$, by the value (\ref{A(1)}) of $A(1)$. We note that in equation (\ref{recurrA4}), $L$, which was initially the size of the system and the degree of the polynomial $R$, no longer needs being an integer. It can assume any complex value such that there is a constant $b$ which solves $A(1)=1$, that is for which the coefficient of $b$ in equation (\ref{recurrA4}) is nonzero when $t=1$. Thus, $L$ must be such that
\begin{equation}
\sum_{j=0}^{n-1}\C{L}{j}(-1)^{j}=(-1)^{n-1}\C{L-1}{n-1}\neq 0\;.
\end{equation}
We find that (\ref{A(1)}) can not be ensured if $L$ is an integer between $1$ and $n-1$, which never happens for the ASEP because of the exclusion rule.\\\indent
We note that the recurrence (\ref{recurrA4}) is singular when $x\to 1$ since $\Delta_{x}$ goes to $0$ in this limit. More precisely, starting from (\ref{A(gamma=0)}), the recurrence equation (\ref{recurrA4}) gives for $A_{1}(t)$ an expression which is regular when $x\to 1$, as $\Delta_{x}^{-1}$ is applied on an a constant independent of $t$ at this order. Using again (\ref{recurrA4}) to calculate $A_{2}(t)$, we see that the $\Delta_{x}^{-1}$ contributes a pole of order $1$ in $x=1$. Iterating (\ref{recurrA4}), we thus see that the $A_{k}(t)$ have a pole of order $k-1$ in $x=1$. As we are interested here in a scaling limit for which $x$ goes to $1$ as the size of the system goes to infinity, this form of the perturbative solution will not be usable. We have to transform it to make it suitable for the weakly asymmetric scaling limit.
\end{section}

\begin{section}{Regularization of the functional equation}
\label{section regularisation}
In this section, we regularize the function $A(t)$ in the limit $x\to 1$ and show that the regularized function can be calculated perturbatively, in a similar way to what we did for $A(t)$ in a previous section.\\\indent
Equations (\ref{A(1)}) and (\ref{E[A]}) indicate that $A(1)$ is regular in $x=1$ and that $A'(1)$ has only a pole of order $1$ in $x=1$. But we also know that at order $k$ in $\gamma$ the function $A(t)$ has a pole of order $k-1$ in $x=1$. This suggests that an expansion in $t=1$ should allow us to regularize the functional equation (\ref{R[A]}) in the limit $x\to 1$. We define
\begin{equation}
\label{Atilde[A]}
\tilde{A}(y)=A(1-(1-x)y)\;.
\end{equation}
In appendix \ref{appendix Atilde regular}, we argue that $\tilde{A}(y)$ is regular when $x\to1$ and work out explicitly how the cancellation of divergent terms in $x=1$ works in the case of a system with only one particle. In the rest of this section, we will write a recurrence equation for $\tilde{A}(y)$ similar to (\ref{recurrA4}) for $A(t)$ but regular in the limit $x\to 1$. This will allow us to study the weakly asymmetric scaling limit. A difference will be that we will now have to do expansion both in $\gamma$ and $1-x$.

\begin{subsection}{Recurrence equation for $\tilde{A}(y)$}
In terms of $\tilde{A}(y)$, the functional Bethe equation (\ref{R[A]}) rewrites
\begin{equation}
\label{R[Atilde]}
\frac{e^{-n\gamma}R(1-(1-x)y)}{(1-x)^{L}}=(1+xy)^{L}\tilde{A}(y)+y^{L}\frac{x^{n}e^{(L-2n)\gamma}}{\tilde{A}(1+xy)}\;,
\end{equation}
while equations (\ref{A(gamma=0)}), (\ref{A(1)}) and (\ref{E[A]}) become respectively
\begin{align}
\label{Atilde(gamma=0)}
\tilde{A}(y)&=1+\O{\gamma}\\
\label{Atilde(0)}
\tilde{A}(0)&=1\\
\label{E[Atilde]}
\frac{E(\gamma,x)}{p}&=\tilde{A}'(0)\;.
\end{align}
We will now solve the functional equation (\ref{R[Atilde]}) perturbatively near $\gamma=0$ and $x=1$. We first write the expansion of $\tilde{A}(y)$ near $\gamma=0$ and $x=1$ as
\begin{equation}
\label{Atildekl}
\tilde{A}(y)=1+\sum_{k=1}^{\infty}\sum_{l=0}^{\infty}\tilde{A}_{k,l}(y)\gamma^{k}(1-x)^{l}\;.
\end{equation}
The range for the summation over $k$ comes from (\ref{Atilde(gamma=0)}) while the range for the summation over $l$ is a consequence of the regularity in $x=1$ of $\tilde{A}(y)$. Since the $A_{k}(t)$ are polynomials (of degree $kn$) in $1/t$, we see that $\tilde{A}(y)$ has only positive integer powers in $y$ after the expansion near $\gamma=0$ and $x=1$. A crucial point in the following will be that the $\tilde{A}_{k,l}(y)$ are in fact polynomials in $y$ of degree $k+l-1$, as can be seen writing $A_{k}(1-(1-x)y)$ at order $l$ in $1-x$ in the following way:
\begin{equation}
\tilde{A}_{k,l}(y)=[A_{k}(1-(1-x)y)]_{(1-x)^{l}}=\left[\sum_{j=0}^{\infty}(-1)^{j}(1-x)^{j}y^{j}\frac{d^{j}\!A_{k}}{dt^{j}}(1)\right]_{(1-x)^{l}}=\sum_{j=0}^{k+l-1}(-1)^{j}y^{j}\left[\frac{d^{j}\!A_{k}}{dt^{j}}(1)\right]_{(1-x)^{l-j}}
\end{equation}
We used the fact that the $A_{k}(t)$ have a pole of order $k-1$ in $x=1$. We will now eliminate the polynomial $R$ from the functional equation (\ref{R[Atilde]}) in a similar way to what we did in the previous section for the functional equation (\ref{R[A]}). We divide the functional equation (\ref{R[Atilde]}) by $y^{L}(1+xy)^{L}$ and make the expansion $y\to\infty$. Taking (\ref{R(infinity)}) into account, we obtain
\begin{equation}
\label{recurrAtilde1}
\frac{\tilde{A}(y)}{y^{L}}+\frac{1}{(1+xy)^{L}}\frac{x^{n}e^{(L-2n)\gamma}}{\tilde{A}(1+xy)}-\frac{(e^{-n\gamma}+x^{n-L}e^{(L-n)\gamma})}{y^{L}}=\O{\frac{1}{y^{L+1}}}\;,
\end{equation}
which must be understood as: each term of the expansion in power series in $\gamma$ and $1-x$ of the l.h.s. is of order $1/y^{L+1}$ when $y\to\infty$. Once again, we see that the polynomial $R$ has disappeared, leaving us with an equation involving only $\tilde{A}(y)$. Replacing $y$ by $y/x$ and dividing everything by $x^{n}$, equation (\ref{recurrAtilde1}) can be rewritten in the slightly more complicated form
\begin{equation}
\label{recurrAtilde2}
\frac{\tilde{A}(y+1)}{(y+1)^{L}}-\frac{\tilde{A}(y)}{y^{L}}=\frac{x^{L-n}\tilde{A}(y/x)-\tilde{A}(y)}{y^{L}}+\frac{1}{(y+1)^{L}}\left(\tilde{A}(y+1)+\frac{e^{(L-2n)\gamma}}{\tilde{A}(y+1)}-x^{L-n}e^{-n\gamma}-e^{(L-n)\gamma}\right)+\O{\frac{1}{y^{L+1}}}\;.
\end{equation}
Defining the finite difference operator $\Delta$ acting on an arbitrary function $u$ as
\begin{equation}
(\Delta u)(y)=u(y+1)-u(y)\;,
\end{equation}
the functional equation (\ref{recurrAtilde2}) finally becomes
\begin{equation}
\label{recurrAtilde3}
\Delta\left(\frac{\tilde{A}(y)}{y^{L}}\right)=\frac{U(y)}{y^{L}}+\frac{V(y+1)}{(y+1)^{L}}+\O{\frac{1}{y^{L+1}}}\;,
\end{equation}
with
\begin{equation}
\label{U[Atilde]}
U(y)=x^{L-n}\tilde{A}(y/x)-\tilde{A}(y)
\end{equation}
and
\begin{equation}
\label{V[Atilde]}
V(y)=\tilde{A}(y)+\frac{e^{(L-2n)\gamma}}{\tilde{A}(y)}-x^{L-n}e^{-n\gamma}-e^{(L-n)\gamma}\;.
\end{equation}
Similarly to what happened in the recurrence equation (\ref{recurrA2}) for $A(t)$, at order $k$ in $\gamma$ and $l$ in $1-x$, the r.h.s. of (\ref{recurrAtilde3}) depends only on the $\tilde{A}_{i,j}$ with either $i=k$ and $j<l$ ($U(y)$) or $i<k$ and $j<l$ ($V(y)$). Thus, equation (\ref{recurrAtilde3}) provides a solution order by order in $\gamma$ and $1-x$ of $\tilde{A}(y)$.
\end{subsection}

\begin{subsection}{Inversion of the operator $\Delta$}
We see that contrary to what happened in (\ref{recurrA2}), the operator $\Delta$ acting on $\tilde{A}$ in the l.h.s. of (\ref{recurrAtilde3}) does not vanish when $x\to 1$. This is related to the fact that that $\tilde{A}(y)$ is not singular in the limit $x\to 1$. The operator $\Delta$ acts formally as
\begin{equation}
\Delta=e^{D_{y}}-1\quad\text{with}\quad D_{y}=d/dy\;.
\end{equation}
Using the Taylor expansion
\begin{equation}
\label{egf Bernoulli}
\frac{z}{e^{z}-1}=\sum_{j=0}^{\infty}\frac{B_{j}}{j!}\,z^{j}\;,
\end{equation}
where the $B_{j}$ are the Bernoulli numbers, we see that we can invert the operator $\Delta$ in (\ref{recurrAtilde3}) by multiplying both sides by $D_{y}/(e^{D_{y}}-1)$. We obtain
\begin{equation}
\label{recurrAtilde4}
D_{y}\left(\frac{\tilde{A}(y)}{y^{L}}\right)=\frac{D_{y}}{e^{D_{y}}-1}\left(\frac{U(y)}{y^{L}}\right)+\frac{-D_{y}}{e^{-D_{y}}-1}\left(\frac{V(y)}{y^{L}}\right)+\O{\frac{1}{y^{L+1}}}\;.
\end{equation}
The differential operators $D_{y}/(e^{D_{y}}-1)$ and $-D_{y}/(e^{-D_{y}}-1)$ must be understood as the formal series (\ref{egf Bernoulli}) in $D_{y}$. We now define some notations that will be useful in the following. For any function $f$, we write the series expansion of $f(y)$ when $y\to\infty$ as
\begin{equation}
f(y)=\sum_{r=a}^{\infty}[f]_{(r)}\,\frac{1}{y^{r}}\;,
\end{equation}
where $a$ is a (possibly negative) integer. We will note $[f(y)]_{(-)}$ the singular part of $f(y)$ when $y\to\infty$ and $[f(y)]_{(+)}$ the non-singular part of $f(y)$ when $y\to\infty$, that is
\begin{equation}
[f(y)]_{(-)}=\sum_{r=a}^{-1}[f]_{(r)}\,\frac{1}{y^{r}}
\end{equation}
and
\begin{equation}
[f(y)]_{(+)}=\sum_{r=0}^{\infty}[f]_{(r)}\,\frac{1}{y^{r}}\;.
\end{equation}
When the function $f$ depends also on $\gamma$ or $x$, we define $[f(y)]_{(-)}$ and $[f(y)]_{(+)}$ such that all the expansions in powers of $1/y$ when $y$ goes to infinity must be done after the expansion in powers of $\gamma$ and $1-x$. Using (\ref{Atildekl}), we can expand $U(y)$ and $V(y)$ near $\gamma=0$ and $x=1$. Since the $\tilde{A}_{k,l}(y)$ are polynomials in $y$, we see that at each order in $\gamma$ and $1-x$, $U(y)$ and $V(y)$ are also polynomials in $y$. We can write
\begin{equation}
U(y)=\sum_{r=0}^{\infty}[U]_{(r)}y^{r}\quad\text{and}\quad V(y)=\sum_{r=0}^{\infty}[V]_{(r)}y^{r}\;.
\end{equation}
With these notations, we integrate equation (\ref{recurrAtilde4}) as a formal series in $y$ and only keep the divergent powers when $y\to\infty$ (that is, the strictly positive powers in $y$). Taking (\ref{Atilde(0)}) into account, we obtain
\begin{equation}
\label{recurrAtilde5}
[\tilde{A}(y)]_{(-)}=\tilde{A}(y)-1=\sum_{r=0}^{\infty}\left[y^{L}\int dy\left([U]_{(r)}\frac{D_{y}}{e^{D_{y}}-1}y^{r-L}+[V]_{(r)}\frac{-D_{y}}{e^{-D_{y}}-1}y^{r-L}\right)\right]_{(-)}\;.
\end{equation}
We did not add a constant term when doing the integration. This can be understood using the analytic continuation for complex $L$ in our formulas: since the $A_{k,l}(y)$ are polynomials in $y$, they have only positive integer powers in $y$. But adding a constant term when integrating (\ref{recurrAtilde4}) would contribute a $y^{L}$ to $\tilde{A}(y)$. This is not possible as we have seen previously that we could take for $L$ any complex value. Using (\ref{egf Bernoulli}), we rewrite (\ref{recurrAtilde5}) as
\begin{equation}
\label{recurrAtilde6}
\tilde{A}(y)-1=\sum_{j=0}^{\infty}\sum_{r=0}^{\infty}B_{j}\C{r-L}{j}\left[y^{L}\int dy\left([U]_{(r)}+(-1)^{j}[V]_{(r)}\right)y^{r-L-j}\right]_{(-)}\;.
\end{equation}
After calculating the integral and using $\C{-a}{j}=(-1)^{j}\C{a+j-1}{j}$, we find
\begin{equation}
\label{recurrAtilde7}
\tilde{A}(y)-1=-\sum_{r=0}^{\infty}\sum_{j=0}^{r}B_{j}\frac{\C{L+j-r-1}{j}}{L+j-r-1}y^{r+1-j}\left((-1)^{j}[U]_{(r)}+[V]_{(r)}\right)\;.
\end{equation}
We observe that the denominator $L+j-r-1$ in this equation makes the previous equation divergent if $L$ is taken to be an integer. Thus, equation (\ref{recurrAtilde7}) must be understood by an analytic continuation in $L$. The binomial coefficient in (\ref{recurrAtilde7}) is thus a polynomial of degree $j$ in $L$. For $j\neq 0$, we see that the denominator $L+j-r-1$ cancels with the binomial coefficient, giving non singular terms in the limit where $L$ becomes an integer. For $j=0$ though, the denominator $L+j-r-1$ does not cancel with the binomial coefficient. It seems to make it impossible to have a finite limit for the r.h.s. of equation (\ref{recurrAtilde7}) when $L$ tends to an integer. But, as we know that $\tilde{A}(y)$ is analytic in $L$ provided that $L\geq n$, it simply means that the numerator contains factors canceling the non-analyticities in $L$ integer larger than $n$. This is indeed shown by using the fact that the $\tilde{A}_{k,l}(y)$ are polynomials in $y$ of degree $k+l-1$. From the definitions (\ref{U[Atilde]}) and (\ref{V[Atilde]}), this implies that $U(y)$ and $V(y)$ are polynomials of degree $k+l-2$ at order $k$ in $\gamma$ and $l$ in $1-x$. Thus, only the terms with $r\leq k+l-2$ contribute to $\tilde{A}_{k,l}(y)$ in equation (\ref{recurrAtilde7}). If we choose $k$ and $l$ such that $k+l\leq L$, we will only need the terms of the sum over $r$ such that $r\leq L-2$. We observe that for these terms, the denominator $L+j-r-1$ is always nonzero, even for $L$ integer. Thus, using the notation
\begin{equation}
\O{\gamma,1-x}^{L+1}\equiv\sum_{s=0}^{L+1}\O{\gamma^{s}}\O{(1-x)^{L+1-s}}\;,
\end{equation}
we can rewrite (\ref{recurrAtilde7}) as
\begin{equation}
\label{recurrAtilde8}
\tilde{A}(y)-1=-\sum_{r=0}^{L-2}\sum_{j=0}^{r}B_{j}\frac{\C{L+j-r-1}{j}}{L+j-r-1}y^{r+1-j}\left((-1)^{j}[U]_{(r)}+[V]_{(r)}\right)+\O{\gamma,1-x}^{L+1}\;.
\end{equation}
In this last equation, we can take $L$ to be an integer: there are no divergences anymore.
\end{subsection}
\end{section}

\begin{section}{Recurrence relations in the weakly asymmetric scaling limit}
\label{section limite WASEP}
In this section, we take the weakly asymmetric limit ($x=1-\nu/L$, $L\to\infty$) of the recurrence relation (\ref{recurrAtilde8}) for $\tilde{A}(y)$. We write closed equations verified by the leading and next-to-leading expressions in the size of the system of a rescaled version of $\tilde{A}(y)$.\\\indent
From the expression (\ref{V[Atilde]}) for $V(y)$, we see that we will have to take $\gamma$ of order $1/L$ in order to obtain a non trivial expression at finite density $\rho=n/L$. It is useful to consider the function $h(y)$
\begin{equation}
\label{h[Atilde]}
h(y)\equiv h(y;\mu,\nu)\equiv\frac{\tilde{A}\left(Ly;\gamma=\frac{\mu}{L},x=1-\frac{\nu}{L}\right)-1}{L}\;.
\end{equation}
From now on, we will no longer use the variables $\gamma$ and $x$. All the expansions in powers of $\gamma$ and $1-x$ will be replaced with expansions in powers of the rescaled variables $\mu$ and $\nu$. Recalling the fact that the $\tilde{A}_{k,l}(y)$ are polynomials in $y$ of degree $k+l-1$, we obtain from (\ref{Atildekl}) that $h(y)$ is of order $1/L$ when the size of the system goes to infinity. We write
\begin{equation}
h(y)=\frac{h_{0}(y)}{L}+\frac{h_{1}(y)}{L^{2}}+\O{\frac{1}{L^{3}}}\;.
\end{equation}
Equations (\ref{Atilde(gamma=0)}) and (\ref{Atilde(0)}) become
\begin{align}
\label{h(mu=0)}
h(y)&=\O{\mu}\\
\label{h(0)}
h(0)&=0\;,
\end{align}
while from (\ref{E[Atilde]}), the rescaled generating function of the cumulants of the current $\tilde{E}(\mu,\nu)$ is given in terms of the function $h$ by
\begin{equation}
\label{Etilde[h]}
\frac{\tilde{E}(\mu,\nu)}{p}\equiv\lim_{L\to\infty}\frac{1}{p}E\left(\gamma=\frac{\mu}{L};x=1-\frac{\nu}{L}\right)=h'(0)\;.
\end{equation}
The binomial coefficient appearing in (\ref{recurrAtilde8}) has the $L\to\infty$ expansion
\begin{equation}
\label{expansion binome}
\openone_{r\leq L-2}\frac{\C{L+j-r-1}{j}}{L+j-r-1}=\frac{L^{j-1}}{j!}\left(1+\frac{1}{L}\left(\frac{j(j-1)}{2}-(r+1)(j-1)\right)+\O{\frac{1}{L^{2}}}\right)\;,
\end{equation}
where $\openone_{r\leq L-2}$ is equal to $1$ if $r\leq L-2$ and $0$ otherwise. Inserting (\ref{expansion binome}) in (\ref{recurrAtilde8}), the recurrence becomes
\begin{equation}
h(y)=-\frac{1}{L}\left[\sum_{r=0}^{\infty}(Ly)^{r}\sum_{j=0}^{\infty}\frac{B_{j}}{j!}\left(1+\frac{1}{L}\left(\frac{y^{2}}{2}\frac{d^{2}}{dy^{2}}+(r+1)y\frac{d}{dy}\right)+\O{\frac{1}{L^{2}}}\right)\left(\frac{(-1)^{j}[U]_{(r)}+[V]_{(r)}}{y^{j-1}}\right)\right]_{(-)}\;.
\end{equation}
Using (\ref{egf Bernoulli}) to perform the summation over $j$, we find
\begin{equation}
h(y)=\frac{1}{L}\left[\sum_{r=0}^{\infty}\left((Ly)^{r}+\frac{1}{L}\left((Ly)^{r}\frac{y^{2}}{2}\frac{d^{2}}{dy^{2}}+\frac{d}{dy}\left(y(Ly)^{r}\right)y\frac{d}{dy}\right)+\O{\frac{1}{L^{2}}}\right)\left(\frac{[U]_{(r)}}{e^{-1/y}-1}-\frac{[V]_{(r)}}{e^{1/y}-1}\right)\right]_{(-)}\;.
\end{equation}
We can finally perform the summation over $r$ and we obtain
\begin{align}
\label{recurrh}
h(y)=&\frac{1}{L}\left[\frac{U(Ly)}{e^{-1/y}-1}-\frac{V(Ly)}{e^{1/y}-1}\right]_{(-)}\\
&+\frac{1}{L^{2}}\left[\frac{y^{2}}{2}\left(\frac{d^{2}}{dy^{2}}\frac{1}{e^{-1/y}-1}\right)U(Ly)+y\left(\frac{d}{dy}\frac{1}{e^{-1/y}-1}\right)\frac{d}{dy}(yU(Ly))\right]_{(-)}\nonumber\\
&-\frac{1}{L^{2}}\left[\frac{y^{2}}{2}\left(\frac{d^{2}}{dy^{2}}\frac{1}{e^{1/y}-1}\right)V(Ly)+y\left(\frac{d}{dy}\frac{1}{e^{1/y}-1}\right)\frac{d}{dy}(yV(Ly))\right]_{(-)}+\O{\frac{1}{L^{3}}}\;.\nonumber
\end{align}
We recall that, from (\ref{U[Atilde]}), (\ref{V[Atilde]}) and (\ref{h[Atilde]}), $U(Ly)$ and $V(Ly)$ are well defined functions depending on $y$ only through $h(y)$. Equation (\ref{recurrh}) is thus a closed equation verified by $h(y)$. It will allow us to obtain both the leading ($h_{0}(y)$) and next-to-leading ($h_{1}(y)$) terms of $h(y)$ in the size of the system. We now have to expand both $U(Ly)$ and $V(Ly)$ to order two in $1/L$ to obtain an equation for $h_{0}$ and an equation for $h_{1}$. The definition of $U$ (\ref{U[Atilde]}) gives
\begin{equation}
U(Ly)=U^{(0)}(y)+\frac{U^{(11)}(y)h_{1}(y)+U^{(10)}(y)}{L}+\O{\frac{1}{L^{2}}}\;,
\end{equation}
with
\begin{align}
\label{U[h]}
U^{(0)}(y)&=(e^{-(1-\rho)\nu}-1)(1+h_{0}(y))\nonumber\\
U^{(11)}(y)&=(e^{-(1-\rho)\nu}-1)\\
U^{(10)}(y)&=e^{-(1-\rho)\nu}\left(\nu yh_{0}'(y)-\frac{(1-\rho)\nu^{2}}{2}(1+h_{0}(y))\right)\;,\nonumber
\end{align}
where $\rho=n/L$ is the particle density. Similarly, using the definition of $V$ (\ref{V[Atilde]}), we find
\begin{equation}
V(Ly)=X[Lh(y)]-x^{L-n}e^{-n\mu}-e^{(L-n)\mu}=V^{(0)}(y)+\frac{V^{(11)}(y)h_{1}(y)+V^{(10)}(y)}{L}+\O{\frac{1}{L^{2}}}\;,
\end{equation}
where we have defined
\begin{equation}
\label{X}
X[f]=1+f+\frac{e^{(1-2\rho)\mu}}{1+f}
\end{equation}
and
\begin{align}
\label{V[h]}
V^{(0)}(y)&=X[h_{0}(y)]-e^{-\rho\mu}e^{-(1-\rho)\nu}-e^{(1-\rho)\mu}\nonumber\\
V^{(11)}(y)&=X'[h_{0}(y)]\\
V^{(10)}(y)&=\frac{(1-\rho)\nu^{2}}{2}e^{-\rho\mu}e^{-(1-\rho)\nu}\;.\nonumber
\end{align}
From (\ref{U[h]}) and (\ref{V[h]}), equation (\ref{recurrh}) gives at the leading order in $\frac{1}{L}$ the following closed equation for $h_{0}(y)$:
\begin{equation}
\label{recurrh0[U,V]}
h_{0}(y)=-\left[\frac{e^{1/y}U^{(0)}(y)+V^{(0)}(y)}{e^{1/y}-1}\right]_{(-)}\;.
\end{equation}
In the next section, we will solve this equation to find $h_{0}(y)$ and thus the leading order of the rescaled generating function of the cumulants of the current. At the next-to-leading order $1/L^{2}$, equation (\ref{recurrh}) leads to
\begin{equation}
\label{recurrh1[alpha,beta]}
[(1-\alpha(y))h_{1}(y)]_{(-)}=[\beta(y)]_{(-)}\;,
\end{equation}
with
\begin{equation}
\label{alpha[U,V]}
\alpha(y)=\frac{U^{(11)}(y)}{e^{-1/y}-1}-\frac{V^{(11)}(y)}{e^{1/y}-1}
\end{equation}
and
\begin{align}
\label{beta[U,V]}
\beta(y)=\frac{U^{(10)}(y)}{e^{-1/y}-1}-\frac{V^{(10)}(y)}{e^{1/y}-1}&+\frac{y^{2}}{2}\left(\frac{d^{2}}{dy^{2}}\frac{1}{e^{-1/y}-1}\right)U^{(0)}(y)+y\left(\frac{d}{dy}\frac{1}{e^{-1/y}-1}\right)\frac{d}{dy}(yU^{(0)}(y))\\
&-\frac{y^{2}}{2}\left(\frac{d^{2}}{dy^{2}}\frac{1}{e^{1/y}-1}\right)V^{(0)}(y)-y\left(\frac{d}{dy}\frac{1}{e^{1/y}-1}\right)\frac{d}{dy}(yV^{(0)}(y))\;.\nonumber
\end{align}
We used the fact that $h_{1}(y)=[h_{1}(y)]_{(-)}$ since $h(0)=0$ (\ref{h(0)}). Once we know $h_{0}(y)$ by solving equation (\ref{recurrh0[U,V]}), equation (\ref{recurrh1[alpha,beta]}) becomes a closed equation for $h_{1}(y)$.
\end{section}

\begin{section}{Solution of the weakly asymmetric functional relations}
\label{section solution WASEP}
In this section, we explicitly solve the functional relations (\ref{recurrh0[U,V]}) and (\ref{recurrh1[alpha,beta]}) to all order in $\mu$ and $\nu$. We obtain the leading and the next-to-leading order in $L$ of the rescaled generating function $\tilde{E}(\mu,\nu)$.

\begin{subsection}{Leading order}
Inserting the expressions (\ref{U[h]}), (\ref{X}) and (\ref{V[h]}) for $U$ and $V$, and using $\left[h_{0}(y)\right]_{(-)}=h_{0}(y)$, the equation (\ref{recurrh0[U,V]}) for the leading order becomes
\begin{equation}
\left(e^{1/y}e^{-(1-\rho)\nu}(1+h_{0}(y))+\frac{e^{(1-2\rho)\mu}}{1+h_{0}(y)}\right)\frac{1}{e^{1/y}-1}=\frac{e^{-\rho\mu}e^{-(1-\rho)\nu}+e^{(1-\rho)\mu}}{e^{1/y}-1}+\O{\frac{1}{y^{0}}}\;.
\end{equation}
Multiplying both sides of the previous equation by $y\left(e^{\frac{1}{2y}}-e^{-\frac{1}{2y}}\right)=1+\O{1/y}$, we obtain
\begin{equation}
\label{recurrh0}
ye^{\frac{1}{2y}}e^{-(1-\rho)\nu}(1+h_{0}(y))+\frac{ye^{-\frac{1}{2y}}e^{(1-2\rho)\mu}}{1+h_{0}(y)}=y\left(e^{-\rho\mu}e^{-(1-\rho)\nu}+e^{(1-\rho)\mu}\right)+\O{\frac{1}{y^{0}}}\;.
\end{equation}
Let us write the unknown function $h_{0}(y)$ as
\begin{equation}
\label{h0(y)}
h_{0}(y)=-1+e^{\frac{(1-2\rho)\mu+(1-\rho)\nu}{2}}e^{\frac{r(y)-1}{2y}}\;.
\end{equation}
We can state (\ref{recurrh0}) as an equation for $r(y)$. We obtain
\begin{equation}
\label{eqr-}
\left[y\cosh\frac{r(y)}{2y}\right]_{(-)}=y\cosh\left(\frac{\mu+(1-\rho)\nu}{2}\right)\;.
\end{equation}
Because of (\ref{h(0)}), we also want $h_{0}(y)$ to have only strictly positive powers in $y$ order by order in $\mu$ and $\nu$. This can be expressed in the form
\begin{equation}
[h_{0}(y)]_{(+)}=0\;.
\end{equation}
In terms of $r(y)$, it becomes
\begin{equation}
\label{eqr+}
\left[e^{\frac{r(y)-1}{2y}}\right]_{(+)}=e^{-\frac{(1-2\rho)\mu+(1-\rho)\nu}{2}}\;.
\end{equation}
We prove in appendix \ref{appendix h0} that
\begin{equation}
\label{r(y)}
r(y)=\sqrt{1-2(1-2\rho)y\mu-2(1-\rho)y\nu+y^{2}(\mu+(1-\rho)\nu)^{2}}
\end{equation}
is the unique formal power series in $\mu$ and $\nu$ that solves (\ref{eqr-}) and (\ref{eqr+}). Thus, equation (\ref{h0(y)}) for $h_{0}(y)$ with $r(y)$ defined by (\ref{r(y)}) is the unique solution of (\ref{recurrh0[U,V]}).
\end{subsection}

\begin{subsection}{Next-to-leading order}
With the expression (\ref{h0(y)}) for $h_{0}(y)$, we can simplify the expressions (\ref{alpha[U,V]}) and (\ref{beta[U,V]}) for $\alpha(y)$ and $\beta(y)$. We begin with $\alpha(y)$. Inserting in (\ref{alpha[U,V]}) the expressions (\ref{U[h]}) and (\ref{V[h]}) of $U^{(11)}(y)$ and $V^{(11)}(y)$, we obtain
\begin{equation}
\label{alpha[X,h]}
\alpha(y)=\frac{e^{-(1-\rho)\nu}-1}{e^{-1/y}-1}-\frac{X'[h_{0}(y)]}{e^{1/y}-1}\;.
\end{equation}
The expressions of the operator $X$ (\ref{X}) and of $h_{0}(y)$ (\ref{h0(y)}) give
\begin{equation}
\label{alpha(y)}
1-\alpha(y)=e^{-(1-\rho)\nu}\frac{1-e^{-r(y)/y}}{1-e^{-1/y}}\;.
\end{equation}
We simplify $\beta(y)$ in appendix \ref{appendix alphabeta}. We find
\begin{align}
\label{beta(y)}
\beta(y)=-&\frac{(1-\rho)\nu^{2}e^{-\rho\mu}e^{-(1-\rho)\nu}}{4\sinh\left(\frac{1}{2y}\right)}\\
&-\left(\frac{\nu(1-(1-\rho)y\nu)}{4y}+\frac{(1-y\nu)(1-(1-2\rho)y\mu-(1-\rho)y\nu)}{4y^{2}r(y)}\right)\frac{e^{\frac{(1-2\rho)\mu-(1-\rho)\nu}{2}}e^{\frac{r(y)}{2y}}}{\sinh\left(\frac{1}{2y}\right)}+\O{\frac{1}{y^{0}}}\;.\nonumber
\end{align}
To be able to solve (\ref{recurrh1[alpha,beta]}), we will need to factorize $1-\alpha(y)$ as a product of a function $u(y)$ with only positive (or zero) powers in $y$ and a function $v(y)$ with only negative (or zero) powers in $y$, after the expansion in powers of $\mu$ and $\nu$. A possible factorization is
\begin{equation}
\label{u(y)}
u(y)=e^{-(1-\rho)\nu}r(y)e^{\frac{1-r(y)}{2y}}
\end{equation}
and
\begin{equation}
\label{v(y)}
v(y)=\frac{\sinh\left(\frac{r(y)}{2y}\right)}{r(y)\sinh\left(\frac{1}{2y}\right)}\;.
\end{equation}
After the expansion in powers of $\mu$ and $\nu$, $r(y)$ is equal to $1$ plus strictly positive powers in $y$ (\ref{r(y)}), which shows that $u(y)$ has indeed only nonnegative powers in $y$. On the contrary, the fact that the hyperbolic sinus has only odd powers gives
\begin{equation}
\frac{\sinh\left(\frac{r(y)}{2y}\right)}{r(y)/2y}=\sum_{j=0}^{\infty}\frac{1}{(2j+1)!}\left(\frac{1}{4y^{2}}-\frac{(1-2\rho)\mu}{2y}-\frac{(1-\rho)\nu}{2y}+\frac{(\mu+(1-\rho)\nu)^{2}}{4}\right)^{j}\;.
\end{equation}
Together with $2y\sinh(1/2y)=1+\O{1/y}$, this proves that $v(y)$ has only negative (or zero) powers in $y$. Writing the equation (\ref{recurrh1[alpha,beta]}) for the next-to-leading order of $h(y)$ as
\begin{equation}
(1-\alpha(y))h_{1}(y)=\beta(y)+\O{\frac{1}{y^{0}}}
\end{equation}
and using the factorization of $1-\alpha(y)$, we divide the previous equation by $v(y)$. We obtain
\begin{equation}
u(y)h_{1}(y)=\frac{\beta(y)}{v(y)}+\O{\frac{1}{y^{0}}}\;.
\end{equation}
Noting that $u(y)h_{1}(y)$ has only strictly positive powers in $y$, we can write
\begin{equation}
u(y)h_{1}(y)=\left[u(y)h_{1}(y)\right]_{(-)}=\left[\frac{\beta(y)}{v(y)}\right]_{(-)}\;.
\end{equation}
Dividing by $u(y)$, we finally find an expression for $h_{1}(y)$
\begin{equation}
\label{h1(y)}
h_{1}(y)=\frac{1}{u(y)}\left[\frac{\beta(y)}{v(y)}\right]_{(-)}\;.
\end{equation}
\end{subsection}

\begin{subsection}{Calculation of the generating function of the cumulants of the current}
We write the rescaled generating function of the cumulants of the current as
\begin{equation}
\frac{\tilde{E}(\mu,\nu)}{p}=\frac{\tilde{E}_{1}(\mu,\nu)}{L}+\frac{\tilde{E}_{2}(\mu,\nu)}{L^{2}}+\O{\frac{1}{L^{3}}}\;.
\end{equation}
Since equation (\ref{Etilde[h]}) expresses $\tilde{E}(\mu,\nu)$ in terms of the derivative of $h(y)$ in $y=0$, we need the expansion of $h(y)$ at first order near $y=0$. We also note that the expression (\ref{h1(y)}) for $h_{1}(y)$ involves the singular part (positive powers in $y$) of the expansion in $1/y$ when $y\to\infty$ of a function of $y$. However, this is not a problem: as $h(y)$ is a polynomial in $y$ at each order in $\mu$ and $\nu$, its expansion when $y\to\infty$ has a finite number of terms, which are all positive powers in $y$. Using the value (\ref{h0(y)}) of $h_{0}(y)$, the generating function (\ref{Etilde[h]}) at the leading order in the size of the system  becomes
\begin{equation}
\tilde{E}_{1}(\mu,\nu)=\rho(1-\rho)(\mu^{2}+\mu\nu)\;.
\end{equation}
For the next-to-leading order, we need the expression (\ref{h1(y)}) of $h_{1}(y)$. Using (\ref{u(y)}), we have
\begin{equation}
u(y)=e^{\frac{(1-2\rho)\mu-(1-\rho)\nu}{2}}+\O{y}\;.
\end{equation}
The next-to-leading order of the eigenvalue is then
\begin{equation}
\label{Etilde2[beta]}
\tilde{E}_{2}(\mu,\nu)=\left[e^{\frac{-(1-2\rho)\mu+(1-\rho)\nu}{2}}\frac{\beta(y)}{yv(y)}\right]_{(1/y)^{0}}\;,
\end{equation}
with the notation $[f(y)]_{(1/y)^{0}}$ for the constant term in the expansion in $1/y$, after the expansion in powers of $\mu$ and $\nu$ as usual. The calculation of this expression is done in appendix \ref{appendix Etilde2(gamma,nu)}. We find for the next-to-leading order of the eigenvalue
\begin{equation}
\tilde{E}_{2}(\mu,\nu)=-\frac{\rho(1-\rho)\mu^{2}\nu}{2}+\sum_{k=1}^{\infty}\frac{B_{2k-2}}{k!(k-1)!}\rho^{k}(1-\rho)^{k}(\mu^{2}+\mu\nu)^{k}\;,
\end{equation}
which concludes the proof of equation (\ref{Etilde(mu,nu)}).
\end{subsection}
\end{section}

\begin{section}{Conclusion}
Exact results for the cumulants of the steady state current in the exclusion process on a ring have already been obtained in the past using Bethe Ansatz: in \cite{ADLvW08.1}, all the cumulants have been calculated in the thermodynamic limit for the symmetric exclusion process, while in \cite{PM08.1,P08.1}, finite size expressions were obtained for the three first cumulants in the system with partial asymmetry. In this paper, we calculated all the cumulants of the current when the asymmetry scales as the inverse of the size of the system (weakly asymmetric exclusion process). We obtain for all the cumulants both the leading and next-to-leading contributions in the size of the system (\ref{Etilde(mu,nu)}).\\\indent
In the scaling of a weak asymmetry, it has been pointed out recently \cite{BD05.1} that the system exhibited a non trivial phase diagram, with in particular a phase of weaker asymmetry in which the fluctuations of the current are gaussian, and a phase of stronger asymmetry for which the fluctuations become non gaussian. On our exact formula (\ref{Etilde(mu,nu)}) for the cumulants of the current, we observe that the next-to-leading order develops singularities when the rescaled asymmetry $\nu$ is larger than some critical value $\nu_{c}$, in perfect agreement with what was predicted in \cite{BD05.1} on the basis of the macroscopic fluctuation theory \cite{BDSGJLL01.1,BDSGJLL04.1} which provides a hydrodynamic description for a large class of driven diffusive systems. Moreover, from a numerical solution of the functional Bethe equations for systems up to size $100$, we confirm that the fluctuations of the current become non gaussian if the asymmetry parameter is larger than $\nu_{c}$. This fact can unfortunately not be seen on the exact formula (\ref{Etilde(mu,nu)}) for the generating function of the cumulants as the non gaussianity of the fluctuations of the current at the leading order is not encoded directly in the large system size limit of the cumulants of the current but is hidden in the non perturbative behavior of their generating function. It would be interesting to calculate by Bethe Ansatz the full form of the generating function of the cumulants, including its non perturbative behavior.\\\indent
We observe on the generating function of the cumulants (\ref{Etilde(mu,nu)}) that the weakly asymmetric case is given by a small deformation of the generating function of the symmetric case, even if the asymmetry parameter becomes larger than the critical value $\nu_{c}$. This deformation can be understood as a minimal way to preserve the Gallavotti--Cohen symmetry. If we go further away from the symmetric case, it is known that the cumulants of the current have more complicated expressions. When the asymmetry scales as the inverse of the square root of the size of the system (crossover between the Edwards-Wilkinson and Kardar-Parisi-Zhang universality classes), the three first cumulants are indeed given by multiple integrals \cite{DM97.1,P08.1}. It is still an open question to calculate all the cumulants of the current in this crossover scaling.\\\indent
Our method for solving the Bethe equations of the system is different from the one used in \cite{ADLvW08.1} for the symmetric case. In that article, the authors used directly the expression of the Bethe equations in terms of the Bethe roots. Their method relies on the fact that the Bethe roots accumulate on a curve in the complex plane as the size of the system goes to infinity. For general Bethe equations, it is in general difficult to know what this curve is: it usually requires a numerical resolution of the Bethe equations, which is not always possible since the Bethe equations are highly coupled. The method we use for solving the Bethe equations, in contrast to the one used in \cite{ADLvW08.1}, does not rely on the behavior of the Bethe roots in the large system size limit. Instead, we use the formulation of the Bethe equations as a functional polynomial equation known as Baxter's TQ equation. This equation can be solved, in the case we are studying, by purely algebraic manipulations. It would be interesting to know if such an approach could be used to study the Bethe equations for some other problems. In particular to calculate the fluctuations of the current for the case of the open ASEP \cite{dGE05.1} and the multispecies ASEP \cite{AB00.1} for which the Bethe equations are already known.

\begin{subsection}*{Acknowledgments}
We thank O. Golinelli for useful discussions.
\end{subsection}
\end{section}

\appendix
\begin{section}{Numerical solution of the functional Bethe equation}
\label{appendix numerical}
We used Newton's method to solve the functional Bethe equation starting from the known solution at $\mu=0$. However, because the coefficients of the polynomial $Q(t)$ do not vary slowly with respect to $\gamma$, we were not able to perform our numerical study on the original functional Bethe equation (\ref{ebQR}). Instead, we used an equivalent equation which does not involve $Q(t)$ but only $R(t)$ \cite{B07.1}. This equation can be obtained from (\ref{ebQR}) in the following way: first, we divide (\ref{ebQR}) by $Q(t)$ and obtain
\begin{equation}
\label{R[Q]1}
R(t)=e^{L\gamma}(1-t)^{L}\frac{Q(xt)}{Q(t)}+(1-xt)^{L}x^{n}\frac{Q(t/x)}{Q(t)}\;.
\end{equation}
Then, we replace $t$ in the previous equation by $t/x$. We have
\begin{equation}
\label{R[Q]2}
R(t/x)=e^{L\gamma}(1-t/x)^{L}\frac{Q(t)}{Q(t/x)}+(1-t)^{L}x^{n}\frac{Q(t/x^{2})}{Q(t/x)}\;.
\end{equation}
Multiplying (\ref{R[Q]1}) and (\ref{R[Q]2}), we see that among the four terms coming in the r.h.s., only one is not proportional to $(1-t)^{L}$. Moreover, all the $Q(t)$ and $Q(t/x)$ cancel in this term. This leaves us with the equation
\begin{equation}
\label{ebRR}
R(t)R(t/x)=x^{n}e^{L\gamma}(1-xt)^{L}(1-t/x)^{L}+\O{(1-t)^{L}}\;.
\end{equation}
This equation provides $L$ constraints on the polynomial $R$. Adding the additional equation
\begin{equation}
R(0)=x^{n}+e^{L\gamma}\;,
\end{equation}
which is a consequence of (\ref{ebQR}), or
\begin{equation}
R(1)=e^{n\gamma}(1-x)^{L}\;,
\end{equation}
which is a consequence of (\ref{Q(1)}) and (\ref{ebQR}), we have enough equations to constrain completely the polynomial $R(t)$, which is of degree $L$. It turns out that the coefficients of $R(t)$ vary much more slowly than the coefficients of $Q(t)$, allowing us to perform a numerical calculation of $E(\gamma,x)$ for systems with up to $50$ particles on $100$ sites. We did these calculations keeping $400$ significant digits for the coefficients of $R(t)$. In the end, we obtained a numerical evaluation of $E(\gamma,x)$ which was symmetric through the Gallavotti--Cohen symmetry, which validated our numerical calculation. A plot of the result for $E(\gamma,x)$ is shown in fig. \ref{fig E(gamma,x) numerical}. The results are discussed in section \ref{section discussion WASEP}.
\end{section}

\begin{section}{Regularity of $\tilde{A}(y)$ in $x=1$}
\label{appendix Atilde regular}
In this appendix, we explain why $\tilde{A}(y)$, defined in (\ref{Atilde[A]}), is regular in $x=1$. In the algebraic formulation of the Bethe Ansatz (see \cite{GM06.1}), one defines a transfer matrix $\tau(\lambda)$ which commutes with the Markov matrix for all complex value of the \textit{spectral parameter} $\lambda$. This transfer matrix can be seen as the generating function over the variable $\lambda$ of non local (and non hermitian) generalized quantum hamiltonians \cite{GM07.1}. The Markov matrix is given in terms of the transfer matrix by the relation $M=pe^{\gamma}\tau'(0)\tau^{-1}(0)$. The largest eigenvalue of the transfer matrix $\epsilon(\lambda)$ can be expressed in terms of the Bethe roots $y_{i}$. An expression for $\epsilon(\lambda)$ is given \textit{e.g.} in \cite{GM06.1}, equation (68), for $\gamma=0$ and in terms of the Bethe roots $z_{i}=e^{\gamma}(1-y_{i})/(1-xy_{i})$ (the authors use the convention $p+q=1$ and $p$ and $q$ are exchanged in comparison to our notations; the generalization to $\gamma\neq 0$ is straightforward). Defining the variable $t$ in terms of the spectral parameter $\lambda$ as
\begin{equation}
t=\frac{1-e^{-\gamma}\lambda}{1-xe^{-\gamma}\lambda}
\end{equation}
we find that $\epsilon(\lambda)$ can be expressed in terms of $Q(t)$ and $R(t)$ as
\begin{equation}
\epsilon(\lambda)=e^{(L-n)\gamma}\left(\frac{1-t}{1-xt}\right)^{L}\frac{Q(xt)}{Q(t)}+x^{n}e^{-n\gamma}\frac{Q(t/x)}{Q(t)}=\frac{e^{-n\gamma}R(t)}{(1-xt)^{L}}\;,
\end{equation}
In the same way that the largest eigenvalue of the deformed Markov matrix is well defined for $x=1$, $\epsilon(\lambda)$ is not singular for $x=1$. In particular, its successive derivatives in $\lambda=0$ correspond to the largest eigenvalue of a well defined generalized hamiltonian and must be regular at $x=1$. In terms of $\tilde{A}(y)$, the eigenvalue $\epsilon(\lambda)$ rewrites
\begin{equation}
\label{epsilon[Atilde]}
\epsilon(\lambda)=\tilde{A}(y)+\lambda^{L}\frac{x^{n}e^{-2n\gamma}}{\tilde{A}(1+xy)}\;,
\end{equation}
while the spectral parameter $\lambda$ can be expressed in terms of $y$ as
\begin{equation}
\lambda(y)=\frac{e^{\gamma}y}{1+xy}\;.
\end{equation}
Taking the successive derivatives of $\epsilon(\lambda(y))$ with respect to $y$, we have
\begin{equation}
\frac{d^{k}}{dy^{k}}\epsilon(\lambda(y))=\sum_{j=1}^{k}\frac{(k-1)!}{(j-1)!}\,\C{k}{j}\,\frac{(-x)^{k-j}e^{j\gamma}}{(1+xy)^{j+k}}\,\frac{d^{j}}{d\lambda^{j}}\epsilon(\lambda)\;,
\end{equation}
which can be checked by recursion on $k$. Inserting the expression (\ref{epsilon[Atilde]}) for $\epsilon(\lambda(y))$ in the previous equation, we observe that the term with $\lambda^{L}/\tilde{A}(1+xy)$ does not contribute for $y=0$ if $k<L$. We obtain for the $k$-th derivative of $\tilde{A}(y)$ at $y=0$ ($k<L$)
\begin{equation}
\left(\frac{d^{k}}{dy^{k}}\tilde{A}(y)\right)_{|y=0}=\sum_{j=1}^{k}\frac{(k-1)!}{(j-1)!}\,\C{k}{j}(-x)^{k-j}e^{j\gamma}\left(\frac{d^{j}}{d\lambda^{j}}\epsilon(\lambda)\right)_{|\lambda=0}\;.
\end{equation}
This shows that $\tilde{A}(y)$ is regular in the vicinity of $x=1$ like $\epsilon(\lambda)$, at least up to order $L-1$ in $y$. To confirm this, we calculated $A(t)$ and $\tilde{A}(y)$ using equation (\ref{recurrA4}) for all systems of size $L\leq 15$ and $n\leq L/2$ up to order $8$ in $\gamma$. In all these cases, we verified that $\tilde{A}(y)$ is regular near $x=1$ at all order in $y$. In the rest of this subsection, we write the complete expressions of $Q(t)$, $A(t)$ and $\tilde{A}(y)$ for systems with one particle on a lattice of size $L$. Again, we observe that the expansion of $A(t)$ in powers of $\gamma$ is singular in $x=1$ while the expansion of $\tilde{A}(y)$ is regular. For $n=1$, $Q(t)$ is a polynomial of degree $1$. It can be written as $Q(t)=t+Q(0)$, the constant $Q(0)$ being set using (\ref{Q(1)}). We find
\begin{equation}
Q(t)=t-\frac{1-e^{\gamma}}{x-e^{\gamma}}\;,
\end{equation}
which does not depend on the size of the system: a particle only feels the finiteness of the lattice through its interactions with the other particles. The generating function of the cumulants of the current is given (\ref{E[Q]}) by
\begin{equation}
E(\gamma,x)=(1-e^{-\gamma})(e^{\gamma}-x)\;.
\end{equation}
Using (\ref{A[Q]}), we find for $A(t)$
\begin{equation}
A(t)=\frac{x\left(\frac{t}{x}-\frac{1-e^{\gamma}}{x-e^{\gamma}}\right)}{e^{\gamma}\left(t-\frac{1-e^{\gamma}}{x-e^{\gamma}}\right)}=\frac{(1-x)t-x(1-t)(1-e^{-\gamma})}{(1-x)t-(1-t)(e^{\gamma}-1)}\;.
\end{equation}
The function $A(t)$ has one pole and one zero, which both tend to zero when $\gamma\to 0$. The beginning of its expansion near $\gamma=0$ is
\begin{align}
A(t)=1+&\left(-1+\frac{1}{t}\right)\gamma+\left(\frac{1}{2}+\frac{x-3}{2(1-x)t}+\frac{1}{(1-x)t^{2}}\right)\gamma^{2}\\
&+\left(-\frac{1}{6}+\frac{7-2x+x^{2}}{6(1-x)^{2}t}-\frac{2}{(1-x)^{2}t^{2}}+\frac{1}{(1-x)^{2}t^{3}}\right)\gamma^{3}+\O{\gamma^{4}}\;.\nonumber
\end{align}
This expansion in singular when $x\to 1$. Taking $t=1-(1-x)y$ in the expression for $A(t)$, a factor $1-x$ cancels between the numerator an the denominator, and we obtain for $\tilde{A}(y)$
\begin{equation}
\tilde{A}(y)=\frac{1-(1-x)y-xy(1-e^{-\gamma})}{1-(1-x)y-y(e^{\gamma}-1)}\;.
\end{equation}
Making the expansion in powers of $\gamma$ and $1-x$, we find
\begin{align}
\tilde{A}(y)=1&+\left((1-x)y+(1-x)^{2}y^{2}+(1-x)^{3}y^{3}+\O{(1-x)^{4}}\right)\gamma\\
&+\left(y+(1-x)\left(-\frac{y}{2}+2y^{2}\right)+(1-x)^{2}\left(-\frac{y^{2}}{2}+3y^{3}\right)+(1-x)^{3}\left(-\frac{y^{3}}{2}+4y^{4}\right)+\O{(1-x)^{4}}\right)\gamma^{2}\nonumber\\
&+\left(y^{2}+(1-x)\left(\frac{y}{6}+3y^{3}\right)+(1-x)^{2}\left(\frac{y^{2}}{6}+6y^{4}\right)+(1-x)^{3}\left(\frac{y^{3}}{6}+10y^{5}\right)+\O{(1-x)^{4}}\right)\gamma^{3}+\O{\gamma^{4}}\;.\nonumber
\end{align}
This expression is indeed regular when $x\to 1$.
\end{section}

\begin{section}{Proof of the expression (\ref{r(y)}) for $r_(y)$}
\label{appendix h0}
In this appendix, we prove that the expression (\ref{r(y)}) for $r(y)$ is the unique formal series in $\mu$ and $\nu$ which solves equations (\ref{eqr-}) and (\ref{eqr+}), and such that (\ref{h(mu=0)}) holds. We first check that $r(y)$ given by (\ref{r(y)}) solves equation (\ref{eqr-}) by direct substitution. We have
\begin{equation}
y\cosh\frac{r(y)}{2y}=y\sum_{j=0}^{\infty}\frac{1}{(2j)!}\left(\frac{r(y)}{2y}\right)^{2j}=y\sum_{j=0}^{\infty}\frac{1}{(2j)!}\left(\frac{(\mu+(1-\rho)\nu)^{2}}{4}-\frac{(1-2\rho)\mu+(1-\rho)\nu}{2y}+\frac{1}{4y^{2}}\right)^{j}\;.
\end{equation}
The $\cosh$ has eliminated the square root of $r(y)$. Taking now only the strictly positive powers in $y$, we obtain
\begin{equation}
\left[y\cosh\frac{r(y)}{2y}\right]_{(-)}=y\sum_{j=0}^{\infty}\frac{1}{(2j)!}\left(\frac{(\mu+(1-\rho)\nu)^{2}}{4}\right)^{j}=y\cosh\left(\frac{\mu+(1-\rho)\nu}{2}\right)\;,
\end{equation}
which is equation (\ref{eqr-}). For equation (\ref{eqr+}), we must as usual do the expansion of $r(y)$ in powers of $\mu$ and $\nu$ before the expansion in powers of $1/y$. Thus, we must write
\begin{equation}
\frac{r(y)-1}{2y}=-\frac{(1-2\rho)\mu+(1-\rho)\nu}{2}+\left(\O{\mu^{2}}+\O{\nu^{2}}+\O{\mu}\O{\nu}\right)\sum_{k=0}^{\infty}\sum_{l=0}^{\infty}yP_{k,l}(y)\mu^{k}\nu^{l}\;,
\end{equation}
where the $P_{k,l}(y)$ are polynomials in $y$. Taking the exponential of the last equation and expanding again in powers of $\mu$ and $\nu$ the term with the double sum over $k$ and $l$, we obtain
\begin{equation}
e^{\frac{r(y)-1}{2y}}=e^{-\frac{(1-2\rho)\mu+(1-\rho)\nu}{2}}\left(1+\left(\O{\mu^{2}}+\O{\nu^{2}}+\O{\mu}\O{\nu}\right)\sum_{k=0}^{\infty}\sum_{l=0}^{\infty}yP_{k,l}(y)\mu^{k}\nu^{l}\right)\;.
\end{equation}
Taking the nonpositive powers in $y$ only eliminates the term with the double sum, leaving us with equation (\ref{eqr+}).\\\indent
The unicity is obtained from the following argument: let us assume that there exists another solution $s(y)$ of the equations (\ref{eqr-}) and (\ref{eqr+}) such that $h_{0}(y)=\O{\mu}$. Because of (\ref{h0(y)}), the condition $h_{0}(y)=\O{\mu}$ gives
\begin{equation}
s(y)=1+\O{\mu}+\O{\nu}\;,
\end{equation}
while equation (\ref{eqr-}) gives
\begin{equation}
\left[y\cosh\frac{r(y)}{2y}\right]_{(-)}=\left[y\cosh\frac{s(y)}{2y}\right]_{(-)}
\end{equation}
and equation (\ref{eqr+}) gives
\begin{equation}
\left[e^{\frac{r(y)-1}{2y}}\right]_{(+)}=\left[e^{\frac{s(y)-1}{2y}}\right]_{(+)}\;.
\end{equation}
Expanding the last two equations at power $k$ in $\mu$ and $l$ in $\nu$, we find by recurrence on $k$ and $l$ that $[r(y)]_{(-)}=[s(y)]_{(-)}$ and $[r(y)/y]_{(+)}=[s(y)/y]_{(+)}$ at all order in $\mu$ and $\nu$. Thus, $r(y)$ and $s(y)$ are equal at each order in $\mu$ and $\nu$ which proves unicity.
\end{section}

\begin{section}{Calculation of $\beta(y)$}
\label{appendix alphabeta}
In this appendix, we simplify the expression (\ref{beta[U,V]}) for $\beta(y)$, taking into account the value (\ref{h0(y)}) of $h_{0}(y)$. Using the identity
\begin{equation}
\frac{y^{2}}{2}\left(\frac{d^{2}}{dy^{2}}\frac{1}{e^{\pm 1/y}-1}\right)f(y)+y\left(\frac{d}{dy}\frac{1}{e^{\pm 1/y}-1}\right)\frac{d}{dy}(yf(y))=\frac{y}{2}\left(\frac{d^{2}}{dy^{2}}\frac{y}{e^{\pm 1/y}-1}\right)f(y)+y^{2}\left(\frac{d}{dy}\frac{1}{e^{\pm 1/y}-1}\right)f'(y)
\end{equation}
valid for an arbitrary function $f$, and the expressions (\ref{U[h]}) and (\ref{V[h]}) of $U^{(0)}(y)$, $U^{(10)}(y)$, $V^{(0)}(y)$ and $V^{(10)}(y)$, we rewrite (\ref{beta[U,V]}) as
\begin{align}
\label{beta[X,h]}
\beta(y)=&\frac{e^{-(1-\rho)\nu}\left(\nu yh_{0}'(y)-\frac{(1-\rho)\nu^{2}}{2}(1+h_{0}(y))\right)}{e^{-1/y}-1}-\frac{\frac{(1-\rho)\nu^{2}}{2}e^{-\rho\mu}e^{-(1-\rho)\nu}}{e^{1/y}-1}\\
&+\left(e^{-(1-\rho)\nu}-1\right)\frac{y}{2}\left(\frac{d^{2}}{dy^{2}}\frac{y}{e^{-1/y}-1}\right)(1+h_{0}(y))+\left(e^{-(1-\rho)\nu}-1\right)y^{2}\left(\frac{d}{dy}\frac{1}{e^{-1/y}-1}\right)h_{0}'(y)\nonumber\\
&-\frac{y}{2}\left(\frac{d^{2}}{dy^{2}}\frac{y}{e^{1/y}-1}\right)\left(X[h_{0}(y)]-e^{-\rho\mu}e^{-(1-\rho)\nu}-e^{(1-\rho)\mu}\right)-y^{2}\left(\frac{d}{dy}\frac{1}{e^{1/y}-1}\right)\frac{d}{dy}X[h_{0}(y)]\;.\nonumber
\end{align}
We use the definition (\ref{X}) of the operator $X$ to express the equation (\ref{recurrh0[U,V]}) for $h_{0}(y)$ as
\begin{equation}
yX[h_{0}(y)]=y\left(e^{(1-\rho)\mu}+e^{-\rho\mu}e^{-(1-\rho)\nu}\right)+y\left(1-e^{-(1-\rho)\nu}e^{1/y}\right)(1+h_{0}(y))+\O{\frac{1}{y^{0}}}\;,
\end{equation}
and its derivative
\begin{equation}
y^{2}\frac{d}{dy}X[h_{0}(y)]=y^{2}\frac{d}{dy}\left[\left(1-e^{-(1-\rho)\nu}e^{1/y}\right)(1+h_{0}(y))\right]+\O{\frac{1}{y^{0}}}\;.
\end{equation}
These last two equations allow us to eliminate all the $X$ operators in the expression (\ref{beta[X,h]}) of $\beta(y)$. We obtain
\begin{equation}
\label{beta[h]}
\beta(y)=-\frac{(1-\rho)\nu^{2}}{2(e^{1/y}-1)}e^{-\rho\mu}e^{-(1-\rho)\nu}-\frac{1-(1-\rho)y^{2}\nu^{2}}{2y^{2}(e^{1/y}-1)}e^{-(1-\rho)\nu}e^{1/y}(1+h_{0}(y))+\frac{1-y\nu}{e^{1/y}-1}e^{-(1-\rho)\nu}e^{1/y}h_{0}'(y)+\O{\frac{1}{y^{0}}}\;.
\end{equation}
From the explicit expressions (\ref{h0(y)}) and (\ref{r(y)}) for $h_{0}(y)$ and $r(y)$, the function $h_{0}'(y)$ can be written in terms of $1+h_{0}(y)$ as
\begin{equation}
h_{0}'(y)=\frac{1+h_{0}(y)}{2y^{2}}\left(1-\frac{1-(1-2\rho)y\mu-(1-\rho)y\nu}{r(y)}\right)\;.
\end{equation}
We insert this in equation (\ref{beta[h]}). It leads to
\begin{align}
\beta(y)=-&\frac{(1-\rho)\nu^{2}e^{-\rho\mu}e^{-(1-\rho)\nu}}{2(e^{1/y}-1)}\\
&-\left(\frac{\nu(1-(1-\rho)y\nu)}{2y}+\frac{(1-y\nu)(1-(1-2\rho)y\mu-(1-\rho)y\nu)}{2y^{2}r(y)}\right)\frac{e^{-(1-\rho)\nu}e^{1/y}}{e^{1/y}-1}(1+h_{0}(y))+\O{\frac{1}{y^{0}}}\;.\nonumber
\end{align}
The first term of $\beta(y)$ in the last equation is equal to $-y(1-\rho)\nu^{2}e^{-\rho\mu}e^{-(1-\rho)\nu}/2+\O{1/y^{0}}$. However, it is better at this point to write all the terms of $\beta(y)$ with $1/\sinh(1/2y)$ in factor. Noting that
\begin{equation}
\frac{1}{e^{1/y}-1}=\frac{e^{-\frac{1}{2y}}}{2\sinh\left(\frac{1}{2y}\right)}=\frac{1}{2\sinh\left(\frac{1}{2y}\right)}+\O{\frac{1}{y^{0}}}\;,
\end{equation}
and using the expression (\ref{h0(y)}) for $1+h_{0}(y)$ in terms of $r(y)$, we finally obtain the following expression for $\beta(y)$
\begin{align}
\beta(y)=-&\frac{(1-\rho)\nu^{2}e^{-\rho\mu}e^{-(1-\rho)\nu}}{4\sinh\left(\frac{1}{2y}\right)}\\
&-\left(\frac{\nu(1-(1-\rho)y\nu)}{4y}+\frac{(1-y\nu)(1-(1-2\rho)y\mu-(1-\rho)y\nu)}{4y^{2}r(y)}\right)\frac{e^{\frac{(1-2\rho)\mu-(1-\rho)\nu}{2}}e^{\frac{r(y)}{2y}}}{\sinh\left(\frac{1}{2y}\right)}+\O{\frac{1}{y^{0}}}\;.\nonumber
\end{align}
\end{section}

\begin{section}{Calculation of $\tilde{E}_{2}(\mu,\nu)$}
\label{appendix Etilde2(gamma,nu)}
In this appendix, we calculate the next-to-leading order of the generating function of the cumulants in the weakly asymmetric scaling, starting from (\ref{Etilde2[beta]}). Using (\ref{beta(y)}) and (\ref{v(y)}), we have
\begin{equation}
\label{Etilde2[r]}
\tilde{E}_{2}(\mu,\nu)=\left[-\frac{(1-\rho)\nu^{2}e^{-\frac{\mu+(1-\rho)\nu}{2}}r(y)}{4y\sinh\left(\frac{r(y)}{2y}\right)}+\frac{\nu(1-(1-\rho)y\nu)r(y)}{2y^{2}\left(e^{-\frac{r(y)}{y}}-1\right)}+\frac{(1-y\nu)(1-(1-2\rho)y\mu-(1-\rho)y\nu)}{2y^{3}\left(e^{-\frac{r(y)}{y}}-1\right)}\right]_{(1/y)^{0}}\;.
\end{equation}
The previous expression for $\tilde{E}_{2}(\mu,\nu)$ has three terms, that we will call respectively $\mathcal{A}$, $\mathcal{B}$ and $\mathcal{C}$. We begin with $\mathcal{A}$. Using the expansion
\begin{equation}
\frac{z}{\sinh z}=\sum_{j=0}^{\infty}\frac{2(1-2^{2j-1})B_{2j}}{(2j)!}z^{2j}
\end{equation}
and the fact that
\begin{equation}
\label{r2(y=infinity)}
\left(\frac{r(y)}{y}\right)^{2}=(\mu+(1-\rho)\nu)^{2}+\O{\frac{1}{y}}\;,
\end{equation}
we find
\begin{equation}
\mathcal{A}=-\frac{(1-\rho)\nu^{2}(\mu+(1-\rho)\nu)}{2\left(e^{\mu+(1-\rho)\nu}-1\right)}\;.
\end{equation}
We now calculate $\mathcal{B}$. Using the expansion (\ref{egf Bernoulli}) and recalling that all the odd $B_{j}$ are equal to $0$ except $B_{1}=-1/2$, we see that
\begin{equation}
\mathcal{B}=\frac{(1-\rho)\nu^{2}}{2}\left(\frac{\mu+(1-\rho)\nu}{e^{\mu+(1-\rho)\nu}-1}+\frac{\mu+(1-\rho)\nu}{2}\right)-\left[\frac{\nu(1-(1-\rho)y\nu)r(y)}{4y^{2}}\right]_{(1/y)^{0}}\;.
\end{equation}
We used here (\ref{r2(y=infinity)}) once again. We need the expansion of $r(y)$ when $y\to\infty$ (again, after the expansion near $\mu=0$ and $\nu=0$). At each order in $\mu$ and $\nu$, $r(y)$ is a polynomial in $y$. We have
\begin{equation}
r(y)=1-((1-2\rho)\mu+(1-\rho)\nu)y+2\rho(1-\rho)(\mu^{2}+\mu\nu)y^{2}+\O{y^{3}}\;,
\end{equation}
which is to be understood after the expansion in powers of $\mu$ and $\nu$ as usual. It finally gives
\begin{equation}
\mathcal{B}=\frac{(1-\rho)\nu^{2}(\mu+(1-\rho)\nu)}{2\left(e^{\mu+(1-\rho)\nu}-1\right)}-\frac{\rho(1-\rho)\mu^{2}\nu}{2}\;.
\end{equation}
Thus, $\mathcal{A}$ and $\mathcal{B}$ only contribute $-\rho(1-\rho)\mu^{2}\nu/2$ to the eigenvalue. We will now see that $\mathcal{C}$ has a non-trivial contribution. Defining
\begin{equation}
f(y)=\frac{1}{e^{-\frac{r(y)}{y}}-1}\;,
\end{equation}
the term $\mathcal{C}$ of equation (\ref{Etilde2[r]}) can be written
\begin{equation}
\label{C[f]}
\mathcal{C}=\frac{\nu((1-2\rho)\mu+(1-\rho)\nu)}{2}\left[f(y)\right]_{(y^{1})}-\frac{(1-2\rho)\mu+\nu+(1-\rho)\nu}{2}\left[f(y)\right]_{(y^{2})}+\frac{1}{2}\left[f(y)\right]_{(y^{3})}\;.
\end{equation}
Expanding $f(y)$ in powers of $r(y)$ with (\ref{egf Bernoulli}), and expanding the powers of $r(y)$ in powers of $y$, we obtain
\begin{equation}
f(y)=\sum_{k=0}^{\infty}\sum_{j=0}^{\infty}\sum_{l=0}^{\infty}\frac{B_{k}}{k!}\C{k-\frac{1}{2}}{j}\C{j}{l}(-1)^{k+l-1}2^{l}[(1-2\rho)\mu+(1-\rho)\nu]^{l}[\mu+(1-\rho)\nu]^{2j-2l}y^{2j+1-k-l}\;.
\end{equation}
We take the term $y^{r}$ in the previous equation, setting $l$ to $2j+1-k-r$ provided that it is nonnegative, and we move out of the sum over $k$ the term $k=1$ which is the only odd $k$ such that $B_{k}\neq 0$. We have
\begin{align}
[f(y)]_{(y^{r})}=-\frac{\delta_{r,0}}{2}+\sum_{k=0}^{\infty}\sum_{j=0}^{\infty}&\openone_{j\geq k+\frac{r-1}{2}}\frac{B_{2k}}{(2k)!}\C{k-\frac{1}{2}}{j}\C{j}{2j+1-2k-r}\\
&\times(-1)^{r}2^{2j+1-2k-r}[(1-2\rho)\mu+(1-\rho)\nu]^{2j+1-2k-r}[\mu+(1-\rho)\nu]^{4k+2r-2j-2}\;.\nonumber
\end{align}
For $r\geq 0$, $j\geq k+\frac{r-1}{2}$ implies $j\geq k$ ($j$ integer). Thus
\begin{equation}
\C{k-\frac{1}{2}}{j}=\frac{(-1)^{j+k}(2k)!(2j-2k)!}{2^{2j}j!k!(j-k)!}\;.
\end{equation}
Replacing $j$ by $j+k$, we obtain
\begin{align}
[f(y)]_{(y^{r})}=-\frac{\delta_{r,0}}{2}+\sum_{k=0}^{\infty}\sum_{j=\left\lceil\frac{r-1}{2}\right\rceil}^{\infty}\frac{B_{2k}}{k!k!}&\frac{(-1)^{j+r}}{2^{2k+r-1}}\frac{(2j)!}{(2j+1-r)!}\frac{(k-j)!}{(k+r-j-1)!}\C{k}{j}\\
&\times[(1-2\rho)\mu+(1-\rho)\nu]^{2j+1-r}[\mu+(1-\rho)\nu]^{2k+2r-2j-2}\;,\nonumber
\end{align}
where $\lceil(r-1)/2\rceil$ is the smallest integer larger than $(r-1)/2$. For $r=1$, the previous formula gives, resumming the sum over $j$
\begin{equation}
[f(y)]_{(y^{1})}=-\sum_{k=0}^{\infty}\frac{B_{2k}}{k!k!}\rho^{k}(1-\rho)^{k}(\mu^{2}+\mu\nu)^{k}\;.
\end{equation}
For $r=2$, we obtain
\begin{equation}
[f(y)]_{(y^{2})}=-\sum_{k=0}^{\infty}\frac{B_{2k}}{k!k!}\rho^{k}(1-\rho)^{k}(\mu^{2}+\mu\nu)^{k}[(1-2\rho)\mu+(1-\rho)\nu]\;,
\end{equation}
and for $r=3$, we have
\begin{equation}
[f(y)]_{(y^{3})}=-\sum_{k=0}^{\infty}\frac{B_{2k}}{k!k!}\rho^{k}(1-\rho)^{k}(\mu^{2}+\mu\nu)^{k}[(1-2\rho)\mu+(1-\rho)\nu]^{2}+2\sum_{k=0}^{\infty}\frac{B_{2k}}{k!(k+1)!}\rho^{k+1}(1-\rho)^{k+1}(\mu^{2}+\mu\nu)^{k+1}\;.
\end{equation}
Inserting the last three equations into the expression (\ref{C[f]}) for $\mathcal{C}$, we see that everything cancels except the second term with $r=3$. Thus, we have
\begin{equation}
\mathcal{C}=\sum_{k=1}^{\infty}\frac{B_{2k-2}}{k!(k-1)!}\rho^{k}(1-\rho)^{k}(\mu^{2}+\mu\nu)^{k}\;.
\end{equation}
Gathering everything, we finally obtain
\begin{equation}
\tilde{E}_{2}(\mu,\nu)=-\frac{\rho(1-\rho)\mu^{2}\nu}{2}+\sum_{k=1}^{\infty}\frac{B_{2k-2}}{k!(k-1)!}\rho^{k}(1-\rho)^{k}(\mu^{2}+\mu\nu)^{k}\;.
\end{equation}
This concludes the proof of the next-to-leading order of equation (\ref{Etilde(mu,nu)}).
\end{section}

\end{document}